\documentclass{aa}  
\usepackage{xurl}  
\usepackage[colorlinks=true, citecolor=blue, linkcolor=blue, urlcolor=blue]{hyperref}

\usepackage{longtable}
\usepackage{fontspec}
\usepackage[utf8]{inputenc}
\usepackage[T1]{fontenc}

\usepackage{graphicx}
\usepackage{txfonts}

\hypersetup{
    colorlinks=true,     
    citecolor=blue,     
    linkcolor=blue,     
    urlcolor=blue,      
}

\usepackage{xcolor}

\begin{document}

   \title{CURLING - III. Identifying Candidates of Wide-separation Gravitationally Lensed Quasars from the CatNorth Catalogue}

    \titlerunning{Wide-separation Lensed Quasars from \texttt{CatNorth}}
    \authorrunning{Di Wu et al.}

   \author{Di Wu
          \inst{1,2}
          \and
          Zizhao He\inst{3,4,5}\fnmsep\thanks{zzhe@ncu.edu.cn}
          \and
          Nan Li\inst{1,2,6}\fnmsep\thanks{nan.li@nao.cas.cn}
          \and
          Shenzhe Cui\inst{1,2}
          \and
          Yuming Fu\inst{9,10}
          \and
          Xue-Bing Wu\inst{7,8}
          \and
          Dan Qiu\inst{7}
          \and 
          Shuaiqing Jiang
          \inst{1,2}
          }

   \institute{National Astronomical Observatories, Chinese Academy of Sciences, 20A Datun Road, Chaoyang District, Beijing 100101, China\\
              \email{nan.li@nao.cas.cn}
         \and
             School of Astronomy and Space Science, University of Chinese Academy of Sciences, Beijing 100049, China
        \and
            Department of Physics, Nanchang University, Nanchang, 330031, China\\
            \email{zzhe@ncu.edu.cn}
        \and
            Center for Relativistic Astrophysics and High Energy Physics, Nanchang University, Nanchang, 330031, China
        \and
            Purple Mountain Observatory, Chinese Academy of Sciences, Nanjing, Jiangsu, 210023, China
        \and
            Key Laboratory of Space Astronomy and Technology, National Astronomical Observatories, Chinese Academy of Sciences, 20A Datun Road, Chaoyang District, Beijing 100101, China
        \and
            Kavli Institute for Astronomy and Astrophysics, Peking University, Yi He Yuan Lu 5, Haidian Qu, 100871 Beijing, China
        \and
            Department of Astronomy, School of Physics, Peking University, Beijing 100871, China
        \and
            Leiden Observatory, Leiden University, The Netherlands
        \and
            Kapteyn Astronomical Institute, University of Groningen, P.O. Box 800, 9700 AV Groningen, The Netherlands
            }

   \date{}

\abstract
   {
Wide-separation lensed quasars (WSLQs) represent a special but rare subclass of strongly lensed quasars with multiple images, magnified by massive galaxy cluster lenses, which offer valuable probes for the properties of dark matter haloes and detailed characteristics of quasar host galaxies. However, only $\sim$ 10 WSLQ systems are known so far, limiting the development of relevant investigations.
   }
   {
To enlarge the sample of WSLQs by mining candidates from large-scale sky surveys, we develop a catalogue-based pipeline and apply it to the \texttt{CatNorth}, which is a quasar candidate catalogue with more than 1.5 million candidates constructed from \textit{Gaia} DR3. The \texttt{CatNorth} has a purity of \(\sim90\%\) and a limiting magnitude of \textit{Gaia} G band \(\lesssim21\).
    }
   {
Our pipeline unfolds in three sequential stages. First, to search for groups of quasar candidates with a maximum quasar image separation between 10 and 72 arcsec, we apply a Friends-of-Friends-like algorithm to the HEALPix grids of \texttt{CatNorth} objects using a grid size of 25.6 arcsec. Second, these identified groups undergo an automatic filtering process that assesses the intra-group similarity of photometric colours or spectral information when available. These two steps yield 14 760 quasar candidate groups, while retaining all discoverable previously known WSLQs. Third, a visual inspection (VI), guided primarily by the projected geometry of the quasar images and plausible foreground objects, yields the final candidate sample with a label indicating their quality.
}
   {
We have identified a total of 333 new WSLQ candidates with separations ranging from 10 to 56.8 arcsec. By exploiting the available \textrm{SDSS} DR16/\textrm{DESI} DR1 spectroscopic data, we uncover two novel WSLQ candidate systems, but 331 WSLQ candidates lack sufficient spectral information, comprising 45 Grade-A, 98 Grade-B, and 188 Grade-C systems. In addition, a sample of 29 dual quasar candidates is presented as a by-product. When feasible, we plan to secure follow-up spectroscopy and deeper imaging to confirm WSLQs from the above candidates and proceed with pertinent scientific investigations.}
  {}
   \keywords{Gravitational lensing: strong -- Galaxies: quasars: general -- Galaxies: clusters: general -- Methods: data analysis -- Catalogs}

   \maketitle
%

\section{Introduction}


A strongly lensed quasar is a quasar whose light is magnified, distorted, and multiplied by the gravitational field of a massive object, such as a foreground galaxy or galaxy cluster, situated along the line of sight \citep{1979Natur.279..381W}. Such multiply imaged strongly lensed quasar systems are of high scientific value, which can be used to study the properties of active galactic nuclei (AGN) through microlensing effects caused by stars in the foreground galaxies \citep{Anguita2008, Motta2012, Fian2024} and measure the Hubble constant via time-delay measurements and to explore the properties of dark matter \citep{Oguri2014, Suyu2014, Wong2020, Kochanek2020, Sonnenfeld2021b}.

In particular, as a species of lensed quasar, the wide-separation lensed quasar (WSLQ) whose maximum separation is greater than 10 arcsec \citep{napier2023coollampsvdiscoverycool} holds particular scientific value. These systems are produced by galaxy group or galaxy cluster scale dark matter haloes' gravitational lensing effect, and their wide separations allow observations of the quasar along multiple, wide-separated lines of sight, which facilitates constructing the three-dimensional spatial distribution of outflows of source AGN \citep{2013AJ....145...48M,2014ApJ...794L..20M, Misawa_2016}. The image positions, time delays, and magnification factors of WSLQs can place constraints on the mass distributions of the cluster-scale dark matter haloes \citep{2017ApJ...835....5S, Martinez_2023,napier2023coollampsvdiscoverycool}. Moreover, the wide separations and high magnifications of WSLQs help resolve quasar host galaxies \citep{2017ApJ...845L..14B, cloonan2024coollampsviiiknownwideseparation}.

The number of WSLQs remains limited. To date, approximately 300 lensed quasars have been discovered \citep[see, e.g.][]{lemon2023}, among them 8 are WSLQs \citep{2003Natur.426..810I,2006ApJ...653L..97I,Oguri_2008,2013ApJ...773..146D, Shu_2018,Shu_2019,2021ApJ...921...42S,Martinez_2023,napier2023coollampsvdiscoverycool}. Despite these findings, significant potential remains for further discoveries of WSLQs within currently available datasets, because the observed number remains far below the prediction \citep{2020MNRAS.495.3727R}.

In this work, we construct a WSLQ candidate catalogue by applying the quasar group-finding method of \cite{2023A&A...672A.123H}. The parent sample is \texttt{CatNorth} \citep{2024ApJS..271...54F}, a new high-quality catalogue derived from the \textit{Gaia} DR3 quasar candidates catalogue \citep{2023A&A...674A..41G}. The purity of the \texttt{CatNorth} quasar candidates catalogue is up to 90$\%$, while the primordial \textit{Gaia} DR3 quasar candidate catalogue only has a purity of about $52\%$ \citep{2023A&A...674A...1G}. The higher purity helps reduce the false positive rate of lensed quasar searching.

The selection has three stages. First, we group \texttt{CatNorth} quasar candidates on the sky with a HEALPix-based Friends-of-Friends-like algorithm, clustering objects that are adjacent in projection. Second, we keep only groups whose maximum pairwise separation exceeds 10 arcsec and subject them to an automatic filter that assesses intra-group similarity in photometric colours or, when available, spectroscopic consistency by the spectrum retrieved from several spectrum datasets. Together these steps reduce \texttt{CatNorth} from \(1\,545\,514\) candidates to \(14\,760\) quasar candidate groups. Finally, visual inspection, guided by the image geometry and the presence of plausible foreground deflectors, yields our final sample of WSLQ candidates; objects in this sample lack sufficient spectroscopic information. In addition, we identify two further valuable WSLQ candidates with available spectra. For the final candidate sample, we perform the cross match with three catalogues of a total of about 1.9 million galaxy clusters to find the most likely WSLQ candidates. Besides, we estimated the completeness of our WSLQ candidates catalogue by testing the discoverable rate of the known WSLQs and assessing our searching algorithm by examining the recovery rate of discoverable WSLQs.

In the process of searching for lensed quasars, dual quasars can often be found as a by-product \citep{2025A&A...695A..76H}. We also generate a dual quasar candidates catalogue in this work. Dual quasars refer to physically associated quasar pairs, typically separated by  1 pc to 100 kpc \citep{DeRosa2019}. The corresponding angular separation extends to $\sim$ 12 arcsec at redshift 2 in the $\Lambda$CDM universe. They are often picked up in searches for lensed quasars because the member quasars of dual quasars lie close to one another in angular position on the sky, which is similar to the angular position relation between the member images of multiply lensed quasar images. Dual quasars serve as valuable objects for investigating the galaxy merger process and the properties of supermassive black holes \citep{BoylanKolchin2008, Roedig2014, Romero2016, Martin2018}. However, the number of known dual quasars remains limited. \cite{2025ApJS..281...25P} reported that, at that time, the number of confirmed dual AGN was only $\sim 160$. More recent work, such as \cite{jing2025} (hereafter J25), identified $\sim 900$ spectroscopically confirmed quasar pairs that are potentially tightly bound, selected by requiring the line-of-sight velocity difference $ < 600~\mathrm{km\,s^{-1}}$ and the projected physical separation $ < 110~\mathrm{kpc}$.

The paper is organised as follows. Section \ref{section: data} introduces the datasets we utilised in this work. In Section \ref{section: method}, we detail our quasar group finding algorithm and the further screening method. The results of the WSLQ candidates catalogue and dual quasar candidates catalogue are shown in Section \ref{section: results}. And the Section \ref{section: discussion} and \ref{section: conclusion} include the discussion and the conclusion of this work, respectively. We adopt a $\Lambda$CDM cosmology in which parameters are from Planck 2018 results \citep{2020A&A...641A...6P}. All magnitudes quoted in this paper are in the AB system.

\section{Datasets}\label{section: data}

Our analysis is anchored in \texttt{CatNorth}, whose properties are summarised in Section~\ref{section: CatNorth}. To aid in the selection of high-value lensed quasar candidates and assess the reliability of our algorithm, we further draw on several auxiliary resources, including spectroscopic archives, galaxy cluster catalogues, and discoverable known WSLQs in \texttt{CatNorth}, introduced in Sections~\ref{section: spectrum dataset}, \ref{section: cluster catalog}, and \ref{section: Discoverable known lenses} respectively.

\subsection{CatNorth} \label{section: CatNorth}

Our search for lensed quasar candidates is performed based on the \texttt{CatNorth} \citep{2024ApJS..271...54F}. \texttt{CatNorth} lists \(1\,545\,514\) quasar candidates extracted from the approximately 6.6 million sources in the \textit{Gaia} DR3 quasar candidate catalogue \citep{2023A&A...674A..41G} and raises the purity to about 90 per cent. This catalogue is primarily selected with a machine learning classification model trained on multi-band photometry; the machine-learning-selected candidates are further purified with proper motions. During this process \cite{2024ApJS..271...54F} utilised \(g\), \(r\), \(i\), \(z\) and \(y\) photometry from Pan-STARRS1 \citep{2016arXiv161205560C} together with \(W1\) and \(W2\) photometry from the CatWISE2020 \citep{2021carwise} and incorporated these in the final \texttt{CatNorth} table. \texttt{CatNorth} covers about \(3\pi\) steradians of sky and has a limiting magnitude of  \(\lesssim21\) in the \textit{Gaia} \(G\) band. For each quasar candidate, it also provides a photometric redshift \(z_{\mathrm{ph}}\), derived using an ensemble regression model. Owing to its large area and high purity, \texttt{CatNorth} constitutes a suitable parent sample for the discovery of WSLQs and dual quasars.

\subsection{Spectrum datasets}\label{section: spectrum dataset}

To refine the lensed quasar candidates found by our quasar group finder, we retrieve the relevant optical spectra from two spectroscopic surveys: SDSS Data Release 16 (SDSS DR16; \citealt{2020ApJS..249....3A}) and the DESI Data Release 1 (DESI DR1; \citealt{desicollaboration2025datarelease1dark}).

The Sloan Digital Sky Survey (SDSS; \citealt{2000AJ....120.1579Y,2011AJ....142...72E,2017AJ....154...28B}) is conducted at the Apache Point Observatory (APO). The quasar catalogue of the SDSS sixteenth data release (DR16; \citealt{2020ApJS..250....8L}) compiles all quasar spectra obtained since the inception of SDSS \citep{2002AJ....123.2945R,2013AJ....145...10D,Dawson_2016}, including observations from the Baryon Oscillation Spectroscopic Survey (BOSS; \citealp{2013AJ....145...10D}) and the Extended BOSS (eBOSS; \citealp{Dawson_2016}), and contains a total of \(750\,414\) quasars, of which \(\sim 500\,000\) lie in the redshift range \(0.8 < z < 2.2\) \citep{Dawson_2016}.

The Dark Energy Spectroscopic Instrument (DESI; \citealt{levi2013desiexperimentwhitepapersnowmass,desicollaboration2016desiexperimentisciencetargeting, DESI_Collaboration_2022}) is carried out by the Mayall 4\,m Telescope at Kitt Peak National Observatory (KPNO). The first public data release, DESI~DR1 \citep{desicollaboration2025datarelease1dark}, constitutes a milestone for spectroscopic sky surveys, providing spectra for 1\,647\,484 quasars reaching a maximum redshift of \(z\sim\) 6-7; about 95\% of these quasars lie at redshifts \(z \sim 0\text{-}3\).

For objects with existing spectra in these catalogues, we apply an automatic filter to reject quasar groups whose spectroscopic information is inconsistent with the nature of strongly lensed quasars. By doing so, we flag a subset of high-priority lensed quasar candidates. In addition to lensed quasar candidates,  we yield a catalogue of dual quasar candidates based on quasar groups' projected separation and line-of-sight velocity difference. Details of the filtering are given in Section \ref{section: automatic selection}.

\subsection{Galaxy cluster catalogue}\label{section: cluster catalog}

We cross match WSLQ candidates obtained in this work with several galaxy group and galaxy cluster catalogues. Systems whose image separations exceed 10 arcsec at a typical redshift configuration (lens redshift at 0.5 and source redshift at 2) require a massive deflector of $\sigma_v \gtrsim$ 522$\rm km\,s^{-1}$ under the simple SIS (Singular Isothermal Sphere) assumption, using the relation between velocity dispersion and mass for massive dark matter halo provided by \cite{2008ApJ...672..122E}, we infer that this value corresponds to the $M_{500}$ of $7.87\times10^{13} M_{\odot}$, typically a galaxy group or cluster \citep{2020ApJS..247...12S}; thus the presence of a known cluster near a candidate improves the probability that the system is a real WSLQ.

We employ three galaxy cluster catalogues:
\begin{itemize}
  \item \textsc{WEN\_CAT} — the compilation of \citet{wen2024catalog158millionclusters}: lists \(1\,581\,179\) clusters with \(M_{500}>4.7\times10^{13}\,M_{\odot}\); the median mass is \(7.9\times10^{13}\,M_{\odot}\); redshifts reach \(z\approx1.5\).
  \item \textsc{ZOU\_CAT} — the catalogue from \citet{Zou_2021}: contains \(540\,432\) clusters with a median \(M_{500}=1.23\times10^{14}\,M_{\odot}\) and \(z\le1\).
  \item \textsc{ERO\_CAT} — the eRASS1 Galaxy Groups and Clusters primary catalogue \citep{bulbul2024srgerositaallskysurveycatalog}: comprises \(12\,247\) clusters and groups spanning \(5\times10^{12}\,M_{\odot}<M_{500}<2\times10^{15}\,M_{\odot}\) and extending to \(z=1.32\).
\end{itemize}
Together, \textsc{WEN\_CAT} and \textsc{ZOU\_CAT} comprise close to two million clusters, offering a particularly rich resource for our studies. Roughly \(40\%\) of \textsc{ZOU\_CAT} objects and \(90\%\) of \textsc{ERO\_CAT} objects have counterparts in \textsc{WEN\_CAT}. The \textsc{ERO\_CAT}, constructed from X-ray observations, is a galaxy cluster/group catalogue that can serve as a valuable complement, because X-ray selection is more sensitive to the true three-dimensional mass distribution, whereas optical selection is more easily affected by structures along the line of sight \citep{2002ARA&A..40..539R,2010MNRAS.407...83E,2011ARA&A..49..409A,2018A&A...620A...1M,2021A&A...652A..12K}. These catalogues allow us to mark the lensed quasar candidates that lie near known clusters and to single out high-priority systems. The relevant cross-matching process is presented in Section~\ref{section: lensed quasar candidates without spectrum}.


\subsection{Known WSLQs}\label{section: Discoverable known lenses}

Up to now, 8 WSLQ systems are known; by the fraction of these lenses that appear in the \texttt{CatNorth} and the fraction of discoverable known lenses that successfully pass through our pipeline, we can estimate the completeness of \texttt{CatNorth} and evaluate the reliability of our lensed quasar candidates search algorithm.

\begin{table*}[htbp]
\caption{Eight known wide-separation lensed quasars (WSLQs) and their basic properties. ``sep$_{\max}$'' is the largest image separation within the system. ``N'' is the intrinsic image multiplicity. ``\texttt{CatNorth} imgs'' is the number of counterpart images found in \texttt{CatNorth}. ``Discoverable'' denotes whether the system is discoverable within \texttt{CatNorth} (i.e.\ has $\geq 2$ counterpart images); $^*$ denotes photometric redshift. The column Brightest Image refers to the g magnitude of the brightest image within the systems; the data is from DESI Legacy Imaging Surveys DR9, except that the $^{**}$ denotes the g magnitude data from Pan-STARRS1.}
\label{tb:wslq_known}
\centering
\small
\begin{tabular}{p{0.03\textwidth}@{\hspace{3pt}}
                p{0.19\textwidth}@{\hspace{3pt}}
                p{0.055\textwidth}@{\hspace{3pt}}
                p{0.055\textwidth}@{\hspace{3pt}}
                p{0.055\textwidth}@{\hspace{3pt}}
                p{0.03\textwidth}@{\hspace{3pt}}
                p{0.07\textwidth}@{\hspace{3pt}}
                p{0.10\textwidth}@{\hspace{3pt}}
                p{0.11\textwidth}@{\hspace{3pt}}
                p{0.17\textwidth}}
\hline\hline
No. & System & $z_{\rm s}$ & $z_{\rm l}$ & sep$_{\max}$ (arcsec) & N & \texttt{CatNorth} imgs & Discoverable & Brightest Image (mag) & Notes / Refs \\
\hline
1 & SDSS J1004+4112         & 1.734 & 0.68    & 14.62 & 5 & 2 & Yes & 19.73 & \cite{2003Natur.426..810I, 2005PASJ...57L...7I} \\
2 & SDSS J1029+2623         & 2.197 & 0.596   & 22.5  & 3 & 2 & Yes & 18.85 & \cite{2006ApJ...653L..97I,Oguri_2008} \\
3 & SDSS J1326+4806         & 2.08  & 0.396   & 21.06 & 2 & 2 & Yes & 20.23 & \cite{Shu_2019} \\
4 & GraL J165105.3$-$041725 & 1.451 & 0.591   & 10.1  & 4 & 2 & Yes & 19.6$^{**}$     & \cite{2021ApJ...921...42S} \\
5 & SDSS J2222+2745         & 2.805 & 0.49    & 15.1  & 6 & 0 & No  & 20.65 & \cite{2013ApJ...773..146D} \\
6 & SDSS J0909+4449         & 2.788 & 0.9$^*$ & 13.86 & 3 & 0 & No  & 21.58 & \cite{Shu_2018} \\
7 & COOL J0542$-$2125       & 1.84  & 0.61    & 25.9  & 3 & 1 & No  & 20.68 & Only brightest image in DESI DR9 matched; \cite{Martinez_2023} \\
8 & COOL J0335$-$1927       & 3.27  & 0.4178  & 23.3  & 3 & 0 & No  & 21.37     & \cite{napier2023coollampsvdiscoverycool} \\
\hline
\end{tabular}
\end{table*}

\begin{figure*}
    \centering
    \begin{minipage}[t]{1\linewidth}
        \centering
        \includegraphics[width=1\linewidth]{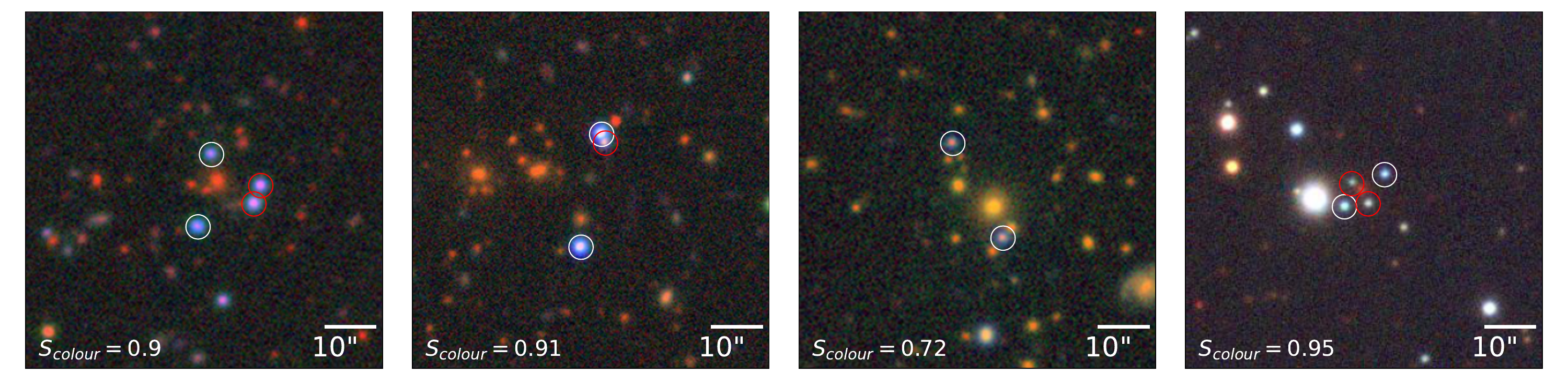}
      
    \end{minipage}

    \caption{ Optical images of the four discoverable wide-separation lensed quasars. From left to right: \textup{SDSS J1004+4112}, \textup{SDSS J1029+2623}, \textup{SDSS J1326+4806}, and \textup{GraL J165105.3$-$041725}. Quasar images included in \texttt{CatNorth} are marked with white circles. The quasar images that belong to known WSLQ systems but are not included in \texttt{CatNorth} are marked as red circles we only plot the non-odd image). The images of the first three panels are from DESI Legacy Imaging Surveys DR9, and the rightmost panel is from Pan-STARRS1. (Orientation: north is up, south is down, west is left, and east is right; same hereafter.)
}
    \label{figure: 4 known lens}
    
\end{figure*}

Among the eight known WSLQs, four are present in \texttt{CatNorth}: J1004+4112 \citep{2003Natur.426..810I,2005PASJ...57L...7I}, SDSS J1029+2623 \citep{2006ApJ...653L..97I, Oguri_2008}, SDSS J1326+4806 \citep{Shu_2019}, and GraL J165105.3-041725 \citep{2021ApJ...921...42S}, i.e., at least two images were found in \texttt{CatNorth}; the remaining WSLQs are missing because they fall below the \texttt{CatNorth}'s limiting magnitude, with only the brightest image of COOL J0542-2125 \citep{Martinez_2023} in the DESI Legacy Imaging Surveys DR9 having a counterpart image in \texttt{CatNorth}. This yields a coverage of approximately 50\% for the known WSLQs within \texttt{CatNorth}. For the four discoverable lenses, \texttt{CatNorth} contains only two images each, their image cutouts are displayed from left to right in Figure \ref{figure: 4 known lens}, while the intrinsic image multiplicities of these systems are 5, 3, 2, and 4 from left to right, respectively. In Section \ref{section: completeness}, we assess the performance of these known lenses in our search pipeline and, on that basis, infer the reliability of our algorithm.

\section{Methodology}\label{section: method}

\begin{figure*}
    \centering
    \begin{minipage}[t]{1\linewidth}
        \centering
        \includegraphics[width=1\linewidth]{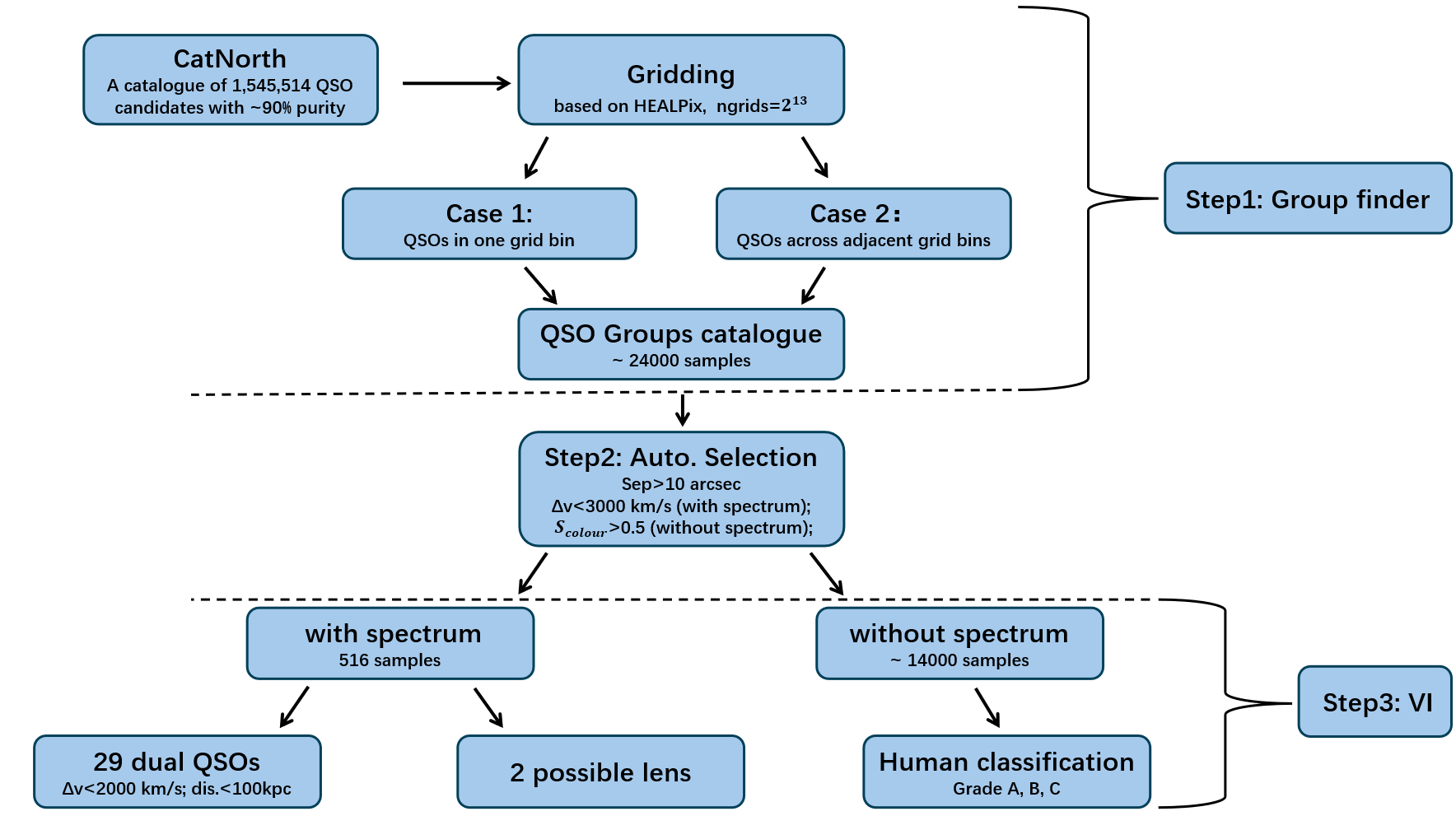}
      
    \end{minipage}

    \caption{Flowchart illustrating the methodology employed to select wide-separation lensed quasar candidates and dual quasar candidates from the \texttt{CatNorth}.}
    \label{process4} 
    
 \end{figure*}

Search for WSLQ and dual quasars in \texttt{CatNorth} is carried out in three stages, summarised in Figure~\ref{process4}. First, we apply the group finder in \citet{2023A&A...672A.123H} to the sky positions of quasar candidates in \texttt{CatNorth} to form a quasar group sample; full details are given in Section~\ref{section: group finder}. Second, an automatic filter based on the spectroscopic or photometric similarity of the members within each group is performed, as described in Section~\ref{section: automatic selection}. In this step, we retain only those quasar groups whose maximum pairwise separation is greater than 10 arcsec, which is the typical scale of a galaxy cluster lensed quasar. Groups in \texttt{CatNorth} with maximum image separation smaller than 10 arcsec are analysed in a companion work (He et al., in prep.). All discoverable known WSLQs present in \texttt{CatNorth} survive these two steps. Third, a visual inspection based on the projected spatial configuration of the group members and the plausible foreground object yields the final samples of lensed quasar candidates and dual quasar candidates, outlined in Section~\ref{section: VI}.

\subsection{Quasar group finder} \label{section: group finder}

We adopt the quasar group search algorithm from \citet{2023A&A...672A.123H} to construct a catalogue of quasar candidate groups whose members' projected angular positions are adjacent on the sky. All \texttt{CatNorth} sources are first assigned to the grid defined by Hierarchical Equal Area isoLatitude Pixelization \citep[HEALPix;][]{1999astro.ph..5275G} with \(N_{\mathrm{side}} = 2^{13}\); this choice gives an angular resolution of about 25.6 arcsec and a total of \(8.1\times10^{8}\) pixels. The 1\,545\,514 objects in \texttt{CatNorth} occupy 1\,542\,794 distinct pixels. For a WSLQ system, two possible cases of its image-pixel mapping are expected. Pixels containing at least two candidates while none of their eight neighbours host any candidates form the Case 1 of the quasar candidates group samples, which comprises 2\,636 groups. Then, starting from every pixel that contains at least one quasar candidate, we examine its surrounding pixels; whenever additional candidates are found, they are iteratively collected, together with their neighbouring pixels, until no further quasar candidates appear. Each assembly of candidates obtained in this way, together with those in the initial pixel, defines a sample labelled Case 2; it contains 20\,096 groups. Combining Case 1 and Case 2 yields 22\,732 quasar candidate groups, which we refer to as the quasar group catalogue (QGC).


Within the QGC, 22 280 groups contain exactly two quasar candidates, and 452 groups contain three or more. For each group with N $\ge$ 3, we also form every subgroup of size 2 to N-1 and add it to the QGC. This adds 1 669 groups, increasing the catalogue to 24 401. The goal is to keep real lenses that might otherwise fail the colour-similarity filter because the original group includes unrelated neighbours with different colours (detailed in Section~\ref{section: automatic selection}).

 \subsection{Automatic screening} \label{section: automatic selection}

From the roughly \(24\,000\) quasar groups returned by the group finder, we conducted a series of automatic screenings.  We first exclude systems whose maximum pairwise separation is $\leq$ 10 arcsec. Then, because the optical spectra and colours of multiple images in a lensed quasar are expected to be highly alike, we then apply an automatic filter based on spectroscopic or photometric similarity to eliminate groups whose members display pronounced discrepancies.

To perform the automatic screening based on the available spectroscopic data, we cross match the positions of member quasar candidates in each group with the SDSS~DR16 and DESI~DR1 within a 1 arcsec radius. We found \(6\,730\) groups possess spectra for at least two members\footnote{When multiple spectra are available for a single member, we preferentially retain the DESI measurement, for whose deeper limiting magnitude usually provides a superior signal-to-noise ratio.}.  For every pair of spectra of member quasar candidates in such a group, we compute the velocity difference by \citep{hogg2000distancemeasurescosmology}:
\begin{equation}
\Delta v = c \, \frac{\Delta z}{1 + z_{\mathrm{mean}}}, 
\label{eq: z2dv}
\end{equation}
where \(\Delta z\) is the redshift difference, \(z_{\mathrm{mean}}\) is the average redshift of the pair, and \(c\) is the speed of light in vacuum.  Groups containing any member pair with \(\Delta v > 3000~\mathrm{km\,s^{-1}}\) are rejected. For the main spectroscopic database used in this work, DESI DR1, the probability of catastrophic spectroscopic redshift failures with velocity errors exceeding \(3000~\mathrm{km\,s^{-1}}\) is extremely low \citep{https://doi.org/10.5281/zenodo.7858207}, such large velocity offsets are highly improbable for multiple images of a lensed quasar.  After this cut, 516 out of 6 730 groups remain.

For groups that lack at least two spectroscopic measurements, we turn to photometry. Following the prescription of \citet{2023A&A...672A.123H}, we calculate the colour similarity \(S_{\mathrm{colour}}\) from the \(g\), \(r\), \(z\), \(W1\), and \(W2\) magnitudes and discard all groups with \(S_{\mathrm{colour}}<0.5\). The definition of colour similarity is:
\begin{equation}
S_{\rm colour} =
\begin{cases}
1 - \dfrac{1}{10}\displaystyle\sum_{i=1}^{10}\sigma_i, & \text{if } \dfrac{1}{10}\displaystyle\sum_{i=1}^{10}\sigma_i < 1,\\[6pt]
0, & \text{if } \dfrac{1}{10}\displaystyle\sum_{i=1}^{10}\sigma_i \ge 1,
\end{cases}
\label{eq:s_colour}
\end{equation}
here, \(N_{\mathrm{colour}}=10\) for the five photometric bands \(\{g,r,z,W1,W2\}\), i.e. the ten colours \(\{g-r,\, g-z,\, g-W1,\, g-W2,\, r-z,\, r-W1,\, r-W2,\, z-W1,\, z-W2,\, W1-W2\}\). For a candidate group with \(n\ge2\) images, \(\sigma_i\) is the standard deviation of the \(i\)-th colour measured across the \(n\) images. All four discoverable known lenses in \texttt{CatNorth} satisfy \(S_{\mathrm{colour}}>0.5\), indicating that this cut preserves real strong lenses. After applying this filter, \(14\,244\) groups without sufficient spectroscopic information remain. We note that, for a direct comparison with the selection criterion of $S_{\mathrm{colour}} > 0.5$ used in \cite{2023A&A...672A.123H}, we choose the same three optical bands, $g$, $r$, and $z$, in our optical-band selection. In this work, applying the selection criterion of $S_{\mathrm{colour}} > 0.5$ removes only $\sim 15\%$ of the groups, which is much lower than the $36\%$ reported in \cite{2023A&A...672A.123H}. This may be because the AGN purity of \texttt{CatNorth}, reaching as high as $\sim 90\%$, makes the original \texttt{CatNorth} sources tend to have a certain degree of colour similarity. In addition, we also tried including the Pan-STARRS1 $i$ band in the calculation of $S_{\mathrm{colour}}$, while still keeping all colours with the same weight and scoring on a scale from 0 to 1, which is similar to Eq. \eqref{eq:s_colour}. We find that the total number of samples selected with the criterion of $S_{\mathrm{colour}} > 0.5$ from the full set of groups found by the group finder differs by only $\sim 4\%$ between the cases with and without the $i$ band, and that the small sub-sample which is only belong to the selection results including the $i$ band all have $S_{\mathrm{colour}} < 0.57$, which is far below that of the known discoverable lenses. This indicates that whether to include the $i$ band has little impact on the total sample size, and has a limited effect on the sample of high-value WSLQ candidates. Therefore, we decided to adopt the same three optical bands, g, r, and z, as in \cite{2023A&A...672A.123H} for our colour similarity calculation.

After the conservative automatic filtering, we retain 516 groups in which at least two quasar candidate images possess optical spectra, together with \(14\,244\) groups that lack adequate optical spectra. We subsequently conduct a visual inspection to isolate systems that are possibly WSLQs.

For the 516 groups with spectra, we first evaluated spectral similarity and the presence or absence of a foreground galaxy cluster. This examination identified two systems that are plausible WSLQs, which are described in Section \ref{section:spectro_cands}; their confirmation or refutation will require higher-quality follow-up observations.  Among the remaining groups, we applied the criteria about velocity difference between group members and the projected distance, subsequently producing 29 dual quasar candidates; the relevant selection criteria and the results are detailed in Section~\ref{section: dual quasar}.

\subsection{Human classification for WSLQs} \label{section: VI}

For the \(14\,244\) groups that lack adequate optical spectra, D.\,W. and S.\,C. carried out a visual inspection (VI) based on their DESI Legacy Imaging Surveys DR9 images \citep{2017PASP..129f4101Z,2019AJ....157..168D} and Pan-STARRS1 images (\citealt{2016arXiv161205560C};  since for a part of the samples, the image from DESI Legacy Imaging Surveys DR9 is not available). Before VI, the inspectors reviewed the DESI Legacy Imaging Surveys DR9 and Pan-STARRS1 images of the four discoverable known lenses to get familiar with the features of true lensed systems. During VI, each one in these \(14\,244\) groups received a score reflecting its likelihood of being a WSLQ, based on three criteria:  
(i) the presence of one or more bright, colour-similar member galaxies of the plausible foreground galaxy cluster near the geometric centre of the quasar images, with preference for a luminous Brightest Cluster Galaxy (BCG);  
(ii) in double-image configurations, the opening angle of the triangle defined by the putative lens and the two images, with larger angles deemed more lens-like;  
(iii) the degree of colour similarity among the quasar images. We note that a relatively large fraction of WSLQs are in the naked cusp configuration (e.g., \citealt{2006ApJ...653L..97I}; \citealt{napier2023coollampsvdiscoverycool}), and in this case the image characteristics do not necessarily satisfy criteria (i), the central lens galaxy in this configuration is not necessarily located between any two images. Therefore, this may cause us to miss lenses of this type in our final results, which is a limitation of our method.

Each group was then graded on a four-point scale (0, 1, 2, 3), where 0 denotes a system that is certainly not a strong lens and higher values indicate increasing probability.  The final score is the average of the two inspectors' assessments.  Groups with a score below 1 were discarded; the remaining systems were retained as WSLQ candidates and assigned quality grades. Following the grading from \cite{2022A&A...662A...4S} and \cite{2025ApJ...981..168H}, we set Grade-C for score = 1, Grade-B for 1 < score $\leq 2$, and Grade-A for score > 2.  This procedure yields 331 candidates, with Grade-C accounting for 57\%, Grade-B 29\%, and Grade-A 14\%, which are described in detail in Section~\ref{section: lensed quasar candidates without spectrum}.

\section{Results}\label{section: results}

We first describe, in Section \ref{section: lensed quasar candidates without spectrum}, the properties of lensed quasar candidates. These objects were selected through photometric colour similarity and visual inspection. The two lensed quasar candidates for which spectra are available are summarised in Section \ref{section:spectro_cands} and are discussed in Appendix \ref{section: appendix}. Section \ref{section: dual quasar} then presents the characteristics of the 29 dual quasar candidates identified in this study.

\subsection{Lensed quasar candidates}\label{section: lensed quasar candidates without spectrum}

We identified a total of 331 lensed quasar candidates (LQC hereafter) without sufficient spectroscopic information, of which 45 are classified as Grade-A, 98 as Grade-B, and 188 as Grade-C. A cross match with existing lensed-quasar candidate catalogues yielded no counterparts, indicating that our candidates are new. These catalogues include: \cite{Dawes2022,2023A&A...672A.123H,chan2023surveygravitationallylensedobjects,lemon2023,Andika_2023,2025ApJ...981..168H,2025A&A...698A..29B}. Among these samples, only \citet{2025A&A...698A..29B} contains systems with separations larger than $10\arcsec$; the other catalogues do not overlap with the LQC because their separations are smaller than $10\arcsec$. \citet{2025A&A...698A..29B} reports a promising candidate that may be a naked cusp lens system or a quadruple-image lens system; this object is not in the LQC but is in the QGC. The region between the two known quasars of this system lacks possible foreground objects, i.e., it does not satisfy criterion (i) in VI, and therefore it was rejected in VI. This also indicates that our working algorithm may miss multiple-image system candidates with $N>2$, such as naked-cusp configurations. Besides, we also checked the quasar pair catalogue J25, and found no overlap with the LQC, because J25 is a sample selected based on DESI DR1 spectroscopy, whereas the LQC is produced from a sample without sufficient DESI DR1 records. To select high-value samples whose projected positions lie near galaxy clusters, a cross-match was carried out between LQC and the three galaxy cluster catalogues (\textsc{WEN\_CAT}, \textsc{ZOU\_CAT}, and \textsc{ERO\_CAT}) using a radius of two arcminutes, because this covers the region where strong lensing by galaxy clusters may occur. The distance is calculated between the centre of the quasar candidates group and the BCG of the galaxy cluster. The relatively large radius is chosen to improve the completeness of the cluster cross match. 108 samples in LQC successfully matched at least one galaxy cluster in these three cluster catalogues, including 21 grade-A, 30 grade-B, and 57 grade-C samples.  The LQC and the cross match results are summarised in a catalogue, which is made available online\footnote{\url{https://github.com/sdwudi/Catalog-of-wide-sep-lensed-QSO-candidates-dual-QSO-from-CatNorth}}.

\begin{figure}
    \centering
    \begin{minipage}[t]{1\linewidth}
        \centering
        \includegraphics[width=1\linewidth]{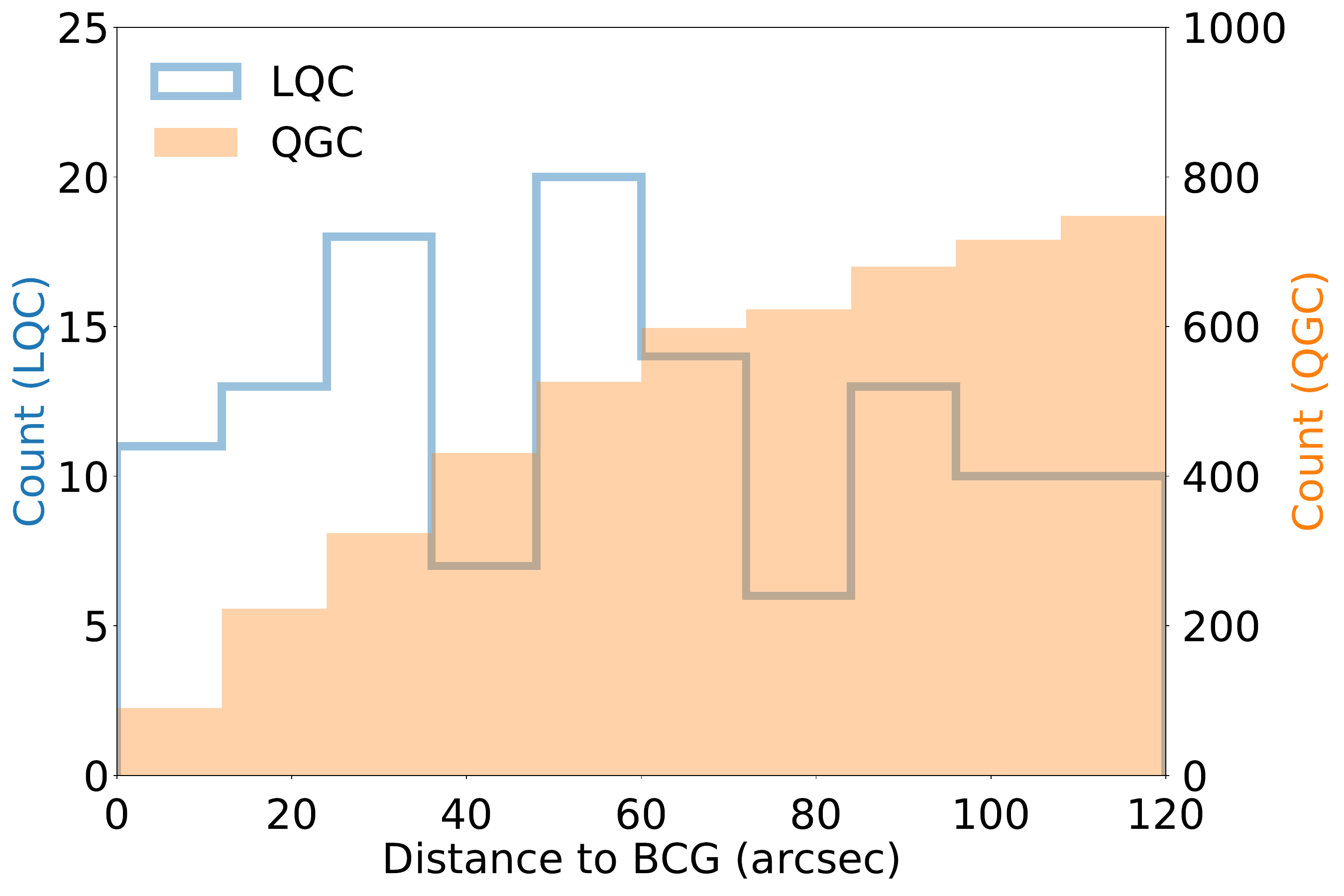}
      
    \end{minipage}

    \caption{\label{figure: BCG_distance_hist} Distribution of the angular separation between the mean sky position of each quasar group and the BCG of the nearest matched cluster (maximum allowed match radius $2\arcmin$). Orange bars (right $y$-axis) show all matches from the QGC (pre-VI), while blue bars (left $y$-axis) show the corresponding results of visually inspected LQC (post-VI).}

\end{figure}

Figure~\ref{figure: BCG_distance_hist} compares the distributions of the angular separation between each quasar group’s mean position and the BCG of the nearest matched cluster for the pre-VI QGC and the post-VI LQC (within a $2\arcmin$ radius). In QGC, the counts grow steadily towards larger radii, a pattern characteristic of chance projections in a fixed search aperture. After visual inspection, LQC shows a clear excess at small separations and a relative deficit at large separations, which is not expected for chance projection. This shift indicates that the visual inspection step preferentially retains systems centred on, or close to, plausible cluster BCGs.

Table~\ref{tb:LQC} lists the columns of our LQC catalogue: A unique groupid (identical for all members belonging to the system), R.A., Dec, VI Grade, maximum separation, number of quasar candidate members within the group, colour similarity, and crossing match results with three galaxy cluster catalogues. In addition to the parameters shown in Table~\ref{tb:LQC}, the online table also retains all columns provided by the original \texttt{CatNorth}.

\begin{figure*}
    \centering
    \begin{minipage}[t]{1\linewidth}
        \centering
        \includegraphics[width=1\linewidth]{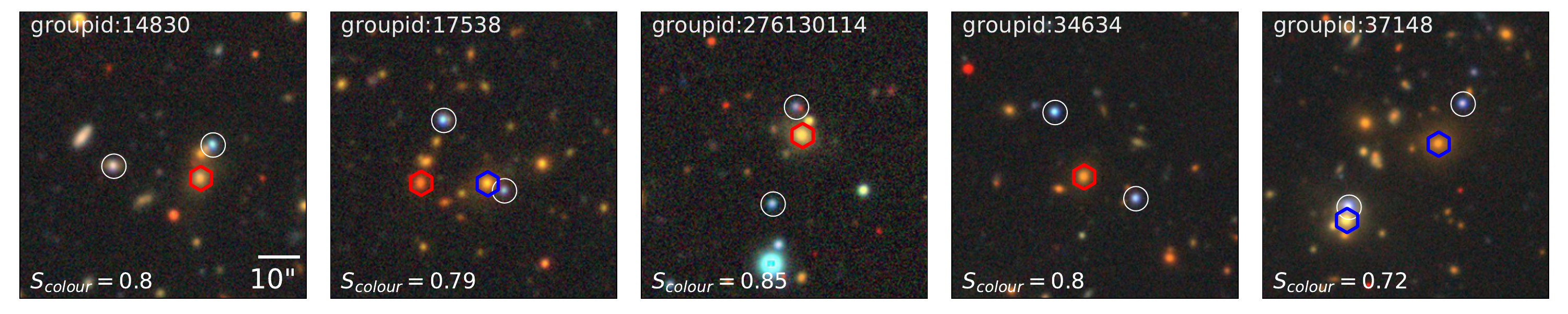}
      
    \end{minipage}
   
    \caption{\label{figure: showcases of wo spec sample with BCG or ero}DESI Legacy Imaging Surveys DR9 \(grz\) composite images of the five Grade-A candidates for which at least one galaxy cluster is located within 30 arcsec of the quasar group centre. Each image spans 70 arcsec $\times$ 70 arcsec. White circles indicate the quasar candidate images, while red and blue circles mark the positions of the BCG matched in \textsc{WEN\_CAT} and \textsc{ZOU\_CAT}, respectively.}
 \end{figure*}

\begin{figure*}
    \centering
    \begin{minipage}[t]{1\linewidth}
        \centering
        \includegraphics[width=1\linewidth]{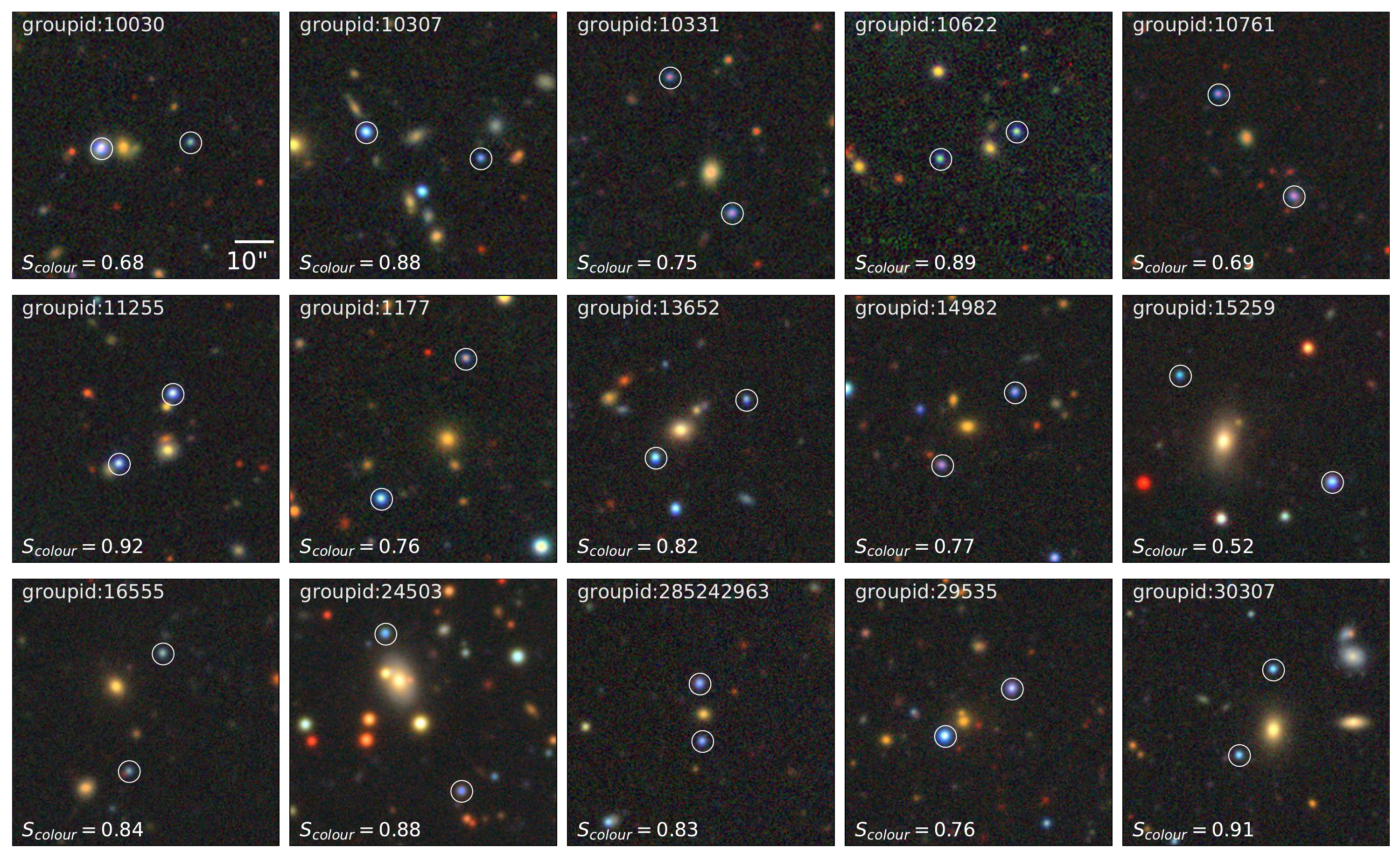}
      
    \end{minipage}
 
    \caption{\label{figure: showcases of wo spec sample without cl}DESI Legacy Imaging Surveys DR9 \(grz\) composite images of the first fifteen Grade-A candidates for which no galaxy cluster is matched within 30 arcsec of the quasar group centre. Each panel spans 70\ arcsec $\times$ 70 arcsec. White circles mark the positions of the quasar candidate images.}
 \end{figure*}

Figure~\ref{figure: showcases of wo spec sample with BCG or ero} displays five Grade-A candidates for which the BCGs of the galaxy cluster are found within 30 arcsec of the mean position of quasar candidate groups. The reason we chose 30 arcsec here is that the core region of the galaxy cluster can provide more matter density to form strong lensing, so that these would be the more promising cases. Besides, Figure~\ref{figure: showcases of wo spec sample without cl} presents images of 15 Grade-A candidates randomly selected from the remainder of the sample, all pictures are drawn from the DESI Legacy Imaging Survey DR9. These examples illustrate that most Grade-A systems exhibit promising lens configurations and therefore constitute the high priority targets for future confirmation.

\begin{figure*}
    \centering
    \begin{minipage}[t]{1\linewidth}
        \centering
        \includegraphics[width=1\linewidth]{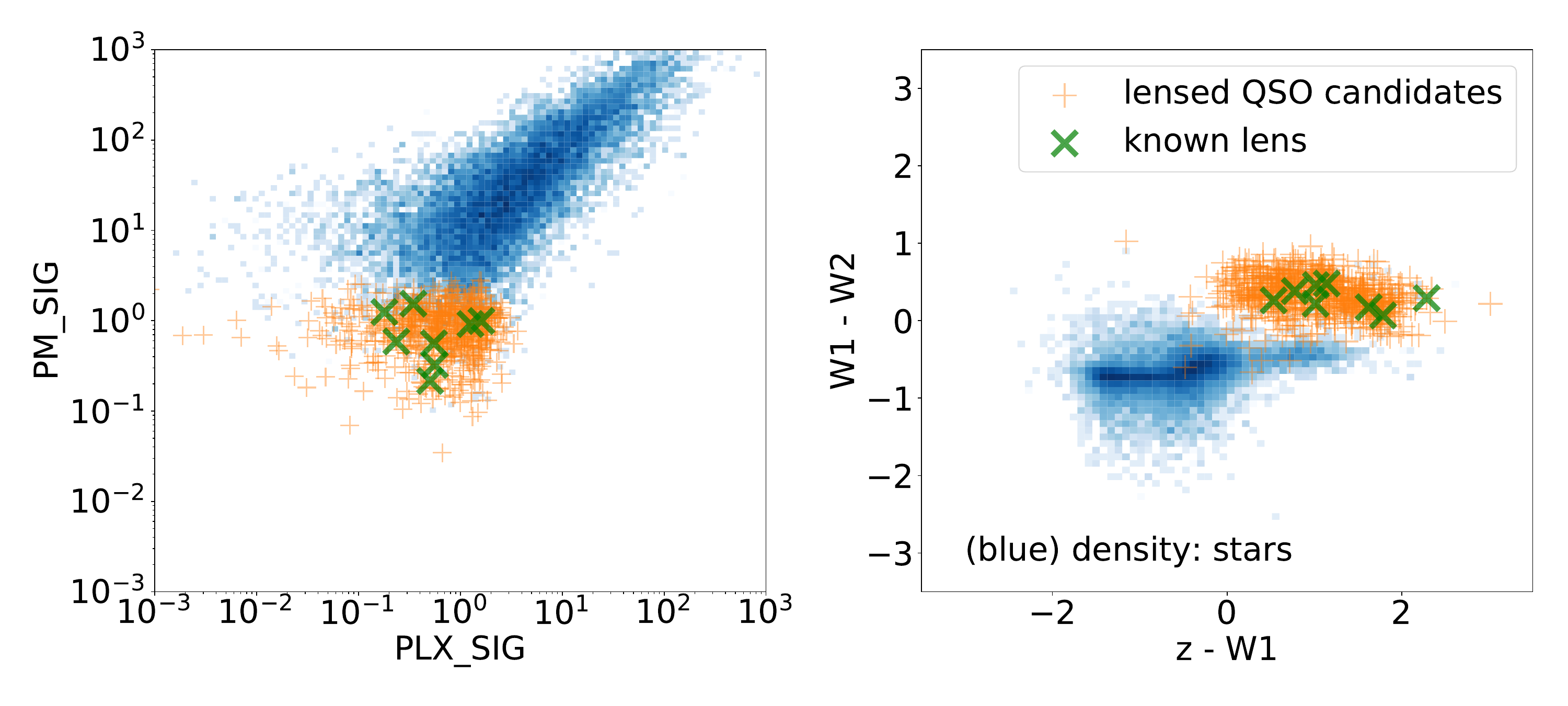}
      
    \end{minipage}

    \caption{\label{figure: pmplx and color of abc candi} The distribution of the properties of the LQC (orange plus signs), the eight images of the four discoverable known lenses in \texttt{CatNorth} (green cross symbols), and the stellar population located within $3\arcmin$ of all LQC objects (blue density map). \textit{Left}: two-dimensional distribution of total $\mathrm{PM\_SIG}$ (the absolute value of the proper motion divided by its uncertainty) versus $\mathrm{PLX\_SIG}$ (the absolute value of the parallax divided by its uncertainty).  \textit{Right}: two-dimensional distribution of the $W1-W2$ and $z-W1$, which is defined in AB magnitude system. We want to note that WISE photometry is calibrated in the Vega system. We convert to AB using $m_{\rm AB}=m_{\rm Vega}+\Delta m$, with $\Delta m_{\rm W1, W2}=(2.699,\,3.339)$, given by \cite{2013wise.rept....1C}.
    }
 \end{figure*}

Figure~\ref{figure: pmplx and color of abc candi} compares the distributions of $\mathrm{PM\_SIG}$, $\mathrm{PLX\_SIG}$, and the colour \(W1-W2\), \(z-W1\) of the LQC, four discoverable WSLQs and a set of stars. $\mathrm{PM\_SIG}\equiv \mu/\sigma_\mu$, where $\mu$ is the total proper motion and $\sigma_\mu$ is its uncertainty (given by \textit{Gaia} DR3; \citealt{2023A&A...674A...1G}), and $\mathrm{PLX\_SIG}\equiv p/\sigma_p$, where $p$ is the parallax and $\sigma_p$ is its uncertainty. The stars are drawn from \textit{Gaia}~DR3 \citep{2023A&A...674A...1G} by requiring (i) an angular distance smaller than \(3\arcmin\) from the quasar candidate group and (ii) a stellar probability \(P_{\mathrm{star}}>0.99\) assigned in the Discrete Source Classifier described by \citet{2023A&A...674A...1G}. By restricting the comparison stars to a small cone around each candidate (here \(<3\arcmin\)) in the first criterion, we ensure similar sky position, which mitigates spatially varying astrometric systematics that depend on position \citep{Lindegren_2018, 2021A&A...649A...1G}.

In Figure \ref{figure: pmplx and color of abc candi}, the candidates occupy similar regions of parameter space as the known lenses, while the stars are clearly segregated. Specifically, in the left panel the candidates cluster at low $\mathrm{PM\_SIG}$ and low $\mathrm{PLX\_SIG}$, as expected for extragalactic sources, whereas the stellar population forms a conspicuous sequence towards large astrometric values. In the right panel, lensed quasar candidates' $(W1-W2,\, z-W1)$ colours trace the discoverable lensed quasar locus and remain well separated from the stellar region. The tight overlap with known lenses and significant separation from stars in these astrometric and colour diagnostics indicate that objects in our sample are high in purity in being real quasars.

\begin{figure*}
    \centering
    \begin{minipage}[t]{1\linewidth}
        \centering
        \includegraphics[width=1\linewidth]{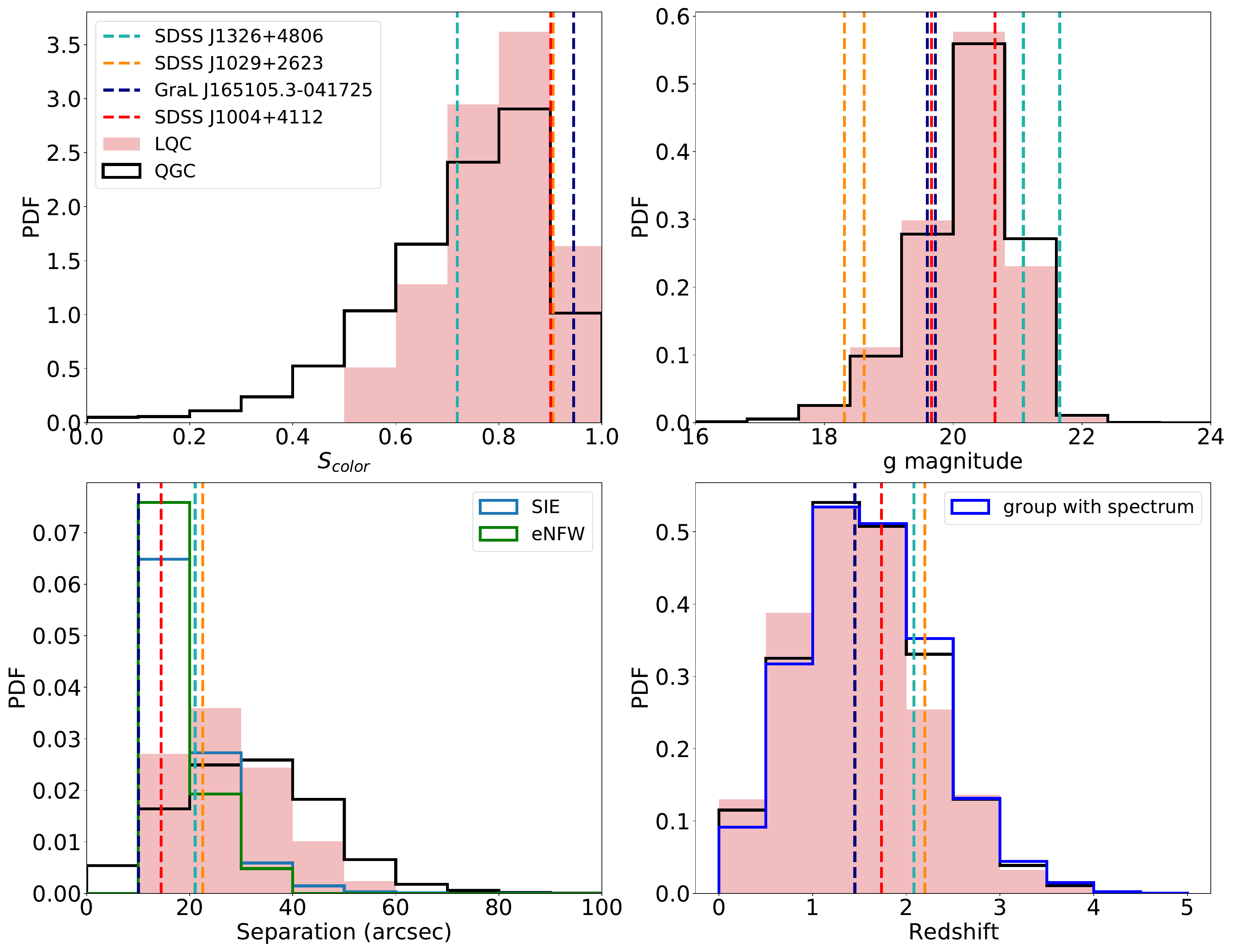}
      
    \end{minipage}

    \caption{\label{figure: wospectrum_stats}Statistical properties of LQC, QGC, and discoverable known lenses.  Upper left: distribution of the colour similarity  \(S_{\mathrm{colour}}\) (we note that the vertical marker line for SDSS~J1004+4112 falls on top of the line for SDSS~J1029+2623, which makes the line for SDSS~J1029+2623 difficult to see).  Upper right: distribution of the Pan-STARRS1 \(g\) band magnitudes for all quasar images contained in each quasar candidate group or known lensed quasar system. Lower left: distribution of the maximum image separation. The blue and green curves are the simulation results derived from the model in Appendix \ref{section: appendixB}, corresponding to different lens mass profile models. Lower right: distribution of the photometric redshift \(z_{\mathrm{ph}}\) for all quasar images in each group or known system; the blue curve shows the \(z_{\mathrm{ph}}\) distribution for those quasar candidate groups in the Quasar Group Catalogue (QGC) that have sufficient spectroscopic matches.  Dashed lines indicate the positions of the corresponding values for the four discoverable known lenses in \texttt{CatNorth}. Black stepped histograms trace the distributions for the QGC.}
\end{figure*}

Figure~\ref{figure: wospectrum_stats} displays the distributions of the \(S_{\mathrm{colour}}\), the Pan-STARRS1 \(g\) band apparent magnitude, the maximum image separation, and the photometric redshift of LQC, QGC, and discoverable quasar images in \texttt{CatNorth} from discoverable known lenses. Vertical dashed lines mark the information of four discoverable known lenses. The \(g\) band magnitude and redshift panels exhibit the information of all member quasar candidate images within each group of LQC and QGC; photometric redshifts are \(z_{\mathrm{ph}}\) provided by \texttt{CatNorth}. Besides, we note that three of four known systems whose Pan-STARRS1 \(g\) band apparent magnitude are plotted in the upper right panel of Figure~\ref{figure: wospectrum_stats} have shown magnitude variations of at least 1 magnitude; these are: SDSS~J1029+2623 \citep{2013ApJ...764..186F}, SDSS~J1326+4806 \citep{Shu_2019}, and SDSS~J1004+4112 \citep{Mu_oz_2022}.

As shown in Figure~\ref{figure: wospectrum_stats}, the \(S_{\mathrm{colour}}\) distribution of LQC shifts towards higher colour similarity compared to QGC. This is due to the removal of groups with low colour similarity during automatic filtering and the favour of colour consistency during visual inspection.

Figure~\ref{figure: wospectrum_stats} also shows the shift of the redshift distribution of LQC towards the low end compared to the original QGC. This has two reasons:  
\begin{itemize}
    \item 6 728 groups rejected by their spectroscopic information tend to reside at higher redshift, because the spectroscopic catalogues employed reach fainter magnitudes than \textit{Gaia}; the redshift distribution information of this part of the samples is plotted as the blue curve in the lower-right panel of Figure~\ref{figure: wospectrum_stats}, so the surviving sample moves to lower redshift; 

    \item during visual inspection, brighter quasars, which are more prevalent at lower redshift, are more readily accepted because they have a higher signal-to-noise ratio, further biasing the accepted sample towards lower redshift.
\end{itemize}

We employed simple models to predict the distribution of the maximum image separation angles of lensed quasars produced by galaxy cluster lenses, in which we adopted the ellipsoidal Navarro-Frenk-White (eNFW) \citep{Navarro_1997, 2002A&A...390..821G} and Singular Isothermal Ellipsoid (SIE) models for the foreground lens. The method to generate this is described in the Appendix \ref{section: appendixB}. The resulting distribution (separation > 10 arcsec) is plotted in the lower left panel of Figure \ref{figure: wospectrum_stats}. From this, we see that the peak of the separation distribution in the LQC sample is located at larger values than predicted by theory: both the SIE and eNFW models peak between 10 and 20 arcsec, whereas the LQC sample peaks between 20 and 30 arcsec. The fractions of systems with separations exceeding 20 arcsec are approximately 25\%, 35\%, and 73\% for eNFW, SIE, and LQC, respectively. These numbers imply that the false positive rate of LQC candidates becomes substantial at the large-separation end.

\begin{table*}[htbp]
\caption{Column description of the WSLQ candidate catalogue (LQC). Note that the units of $M_{\rm 500}$ for each catalogue follow the conventions of the original publications and are therefore not homogenised here.}
\label{tb:LQC}
\centering
\small
\begin{tabular}{p{0.04\textwidth}@{\hspace{3pt}}
                p{0.20\textwidth}@{\hspace{3pt}}
                p{0.06\textwidth}@{\hspace{3pt}}
                p{0.07\textwidth}@{\hspace{3pt}}
                p{0.55\textwidth}}
\hline\hline
Col. & Name & Type & Unit & Description\\
\hline
1  & \texttt{ra}                 & double & deg                     & \textit{Gaia}\,DR3 right ascension (ICRS, epoch 2016.0)\\
2  & \texttt{dec}                & double & deg                     & \textit{Gaia}\,DR3 declination (ICRS, epoch 2016.0)\\
3  & \texttt{groupid}          & int  & ---                     & Candidate lens system identifier\\
4  & \texttt{Grade}              & string & ---                     & System grade (A, B, or C)\\
5  & \texttt{sep\_max}           & float  & arcsec                  & Maximum angular separation between quasar members\\
6  & \texttt{z\_diff}            & float  & ---                     & Maximum $\Delta z_{\mathrm{ph}}$ among lensed quasar candidate system members\\
7  & \texttt{quasar\_num}           & int    & ---                     & Number of quasar members\\
8  & \texttt{S\_colour}          & float  & ---                     & Colour-similarity statistic\\
9  & \texttt{Wen\_BCG\_RA}       & float  & deg                     & R.A.\ of matched BCG in WEN\_CAT\\
10 & \texttt{Wen\_BCG\_DEC}      & float  & deg                     & Dec.\ of the same BCG\\
11 & \texttt{Wen\_redshift}      & float  & ---                     & Cluster redshift in WEN\_CAT\\
12 & \texttt{Wen\_M500}          & float  & $10^{14}$\,M$_\odot$    & $M_{500}$ in WEN\_CAT\\
13 & \texttt{Wen\_ID}            & int    & ---                     & Cluster identifier in WEN\_CAT\\
14 & \texttt{Zou\_BCG\_RA}       & float  & deg                     & R.A.\ of matched BCG in ZOU\_CAT\\
15 & \texttt{Zou\_BCG\_DEC}      & float  & deg                     & Dec.\ of the same BCG\\
16 & \texttt{Zou\_redshift}      & float  & ---                     & Cluster redshift in ZOU\_CAT\\
17 & \texttt{Zou\_M500}          & float  & $\log_{10}$(M$_\odot$)  & $M_{500}$ in ZOU\_CAT\\
18 & \texttt{Zou\_ID}            & int    & ---                     & Cluster identifier in ZOU\_CAT\\
19 & \texttt{eROSITA\_RA}        & float  & deg                     & R.A.\ of matched eROSITA cluster\\
20 & \texttt{eROSITA\_DEC}       & float  & deg                     & Dec.\ of the same cluster\\
21 & \texttt{eROSITA\_redshift}  & float  & ---                     & eROSITA cluster redshift\\
22 & \texttt{eROSITA\_M500}      & float  & $10^{13}$\,M$_\odot$    & $M_{500}$ in eROSITA catalogue\\
23 & \texttt{eROSITA\_ID}        & int    & ---                     & Cluster identifier in eROSITA catalogue\\
\hline
\end{tabular}
\end{table*}

\subsection{Candidates with spectra}\label{section:spectro_cands}

We highlight two high-priority wide-separation candidates for which optical spectra are available (Figures~\ref {figure: 359463625} and \ref{figure: 28001}). These figures are promoted here from the Appendix for ease of reference; full descriptions of the spectral assessments and lens modelling are provided in Appendix~\ref{section: appendix}, $\Delta\theta$ denotes image separation.

\subsubsection{\rm J110121.67+060931.3 ($\Delta\theta=14.14\arcsec$).}
\begin{figure*}
    \centering
    \begin{minipage}[t]{1\linewidth}
        \centering
        \includegraphics[width=1\linewidth]{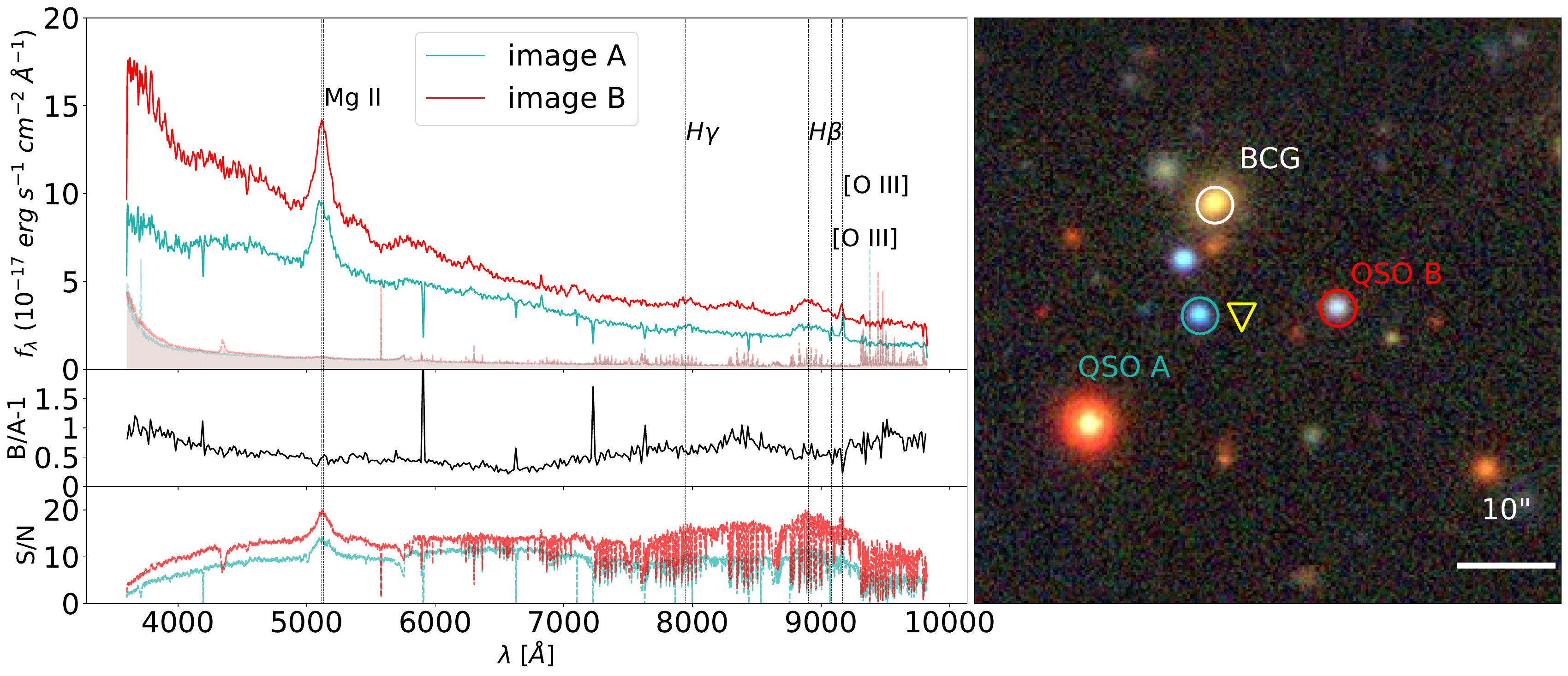}
      
    \end{minipage}
   
    \caption{\label{figure: 359463625}Spectra and image of \textup{J110121.67+060931.3}, the image separation is $14.14\arcsec$. Left: DESI~DR1 spectra of images A and B (smoothed for clarity; shaded bands show the unsmoothed noise); reference lines correspond to $z=0.8305$. Right: DESI Legacy Surveys DR9 image; the white circle marks the BCG from \textsc{WEN\_CAT}, and the yellow triangle the eRASS1 X-ray source. See Appendix~\ref{app_section: samples with spectrumm_1} for details.}
 \end{figure*}

DESI~DR1 spectra of the two images show broadly consistent quasar features near $z\approx0.83$ (Figure \ref{figure: 359463625}). Imaging reveals a plausible foreground cluster: a BCG candidate from \textsc{WEN\_CAT} and an eRASS1 X-ray source \citep{2024A&A...682A..34M} in the field. An SIE mass model centred at the X-ray peak (yellow inverted triangle in the right panel of Figure \ref{figure: 359463625}) reproduces the observed configuration, with a best fit $\sigma_v\simeq608.6~\mathrm{km\,s^{-1}}$, $q\simeq0.89$, $\phi\simeq19.3^\circ$, and an Einstein radius $\theta_{\rm E}\simeq7.19\arcsec$; the enclosed mass is $M(<\theta_{\rm E})\simeq7.4\times10^{12}~M_\odot$.
This is compatible with an X-ray-inferred cluster mass of $M_{500}\gtrsim2\times10^{14}~M_\odot$. We note that this X-ray source is classified as a `point-like' source in the eRASS1 X-ray catalogues \citep{2024A&A...682A..34M}, which makes the assumption that the X-ray emission corresponds to diffuse emission from the galaxy cluster dubious. In addition, we find that this object is also present in the catalogue of J25 and is flagged as a common quasar pair in J25. Further deep imaging and spectroscopy are needed to pin down the mass centroid and test for additional faint images.
(For details, see Appendix~\ref{app_section: samples with spectrumm_1}.)

\subsubsection{\rm J150155.61$-$025728.4 ($\Delta\theta=19.32\arcsec$).}
\begin{figure*}
    \centering
    \begin{minipage}[t]{1\linewidth}
        \centering
        \includegraphics[width=1\linewidth]{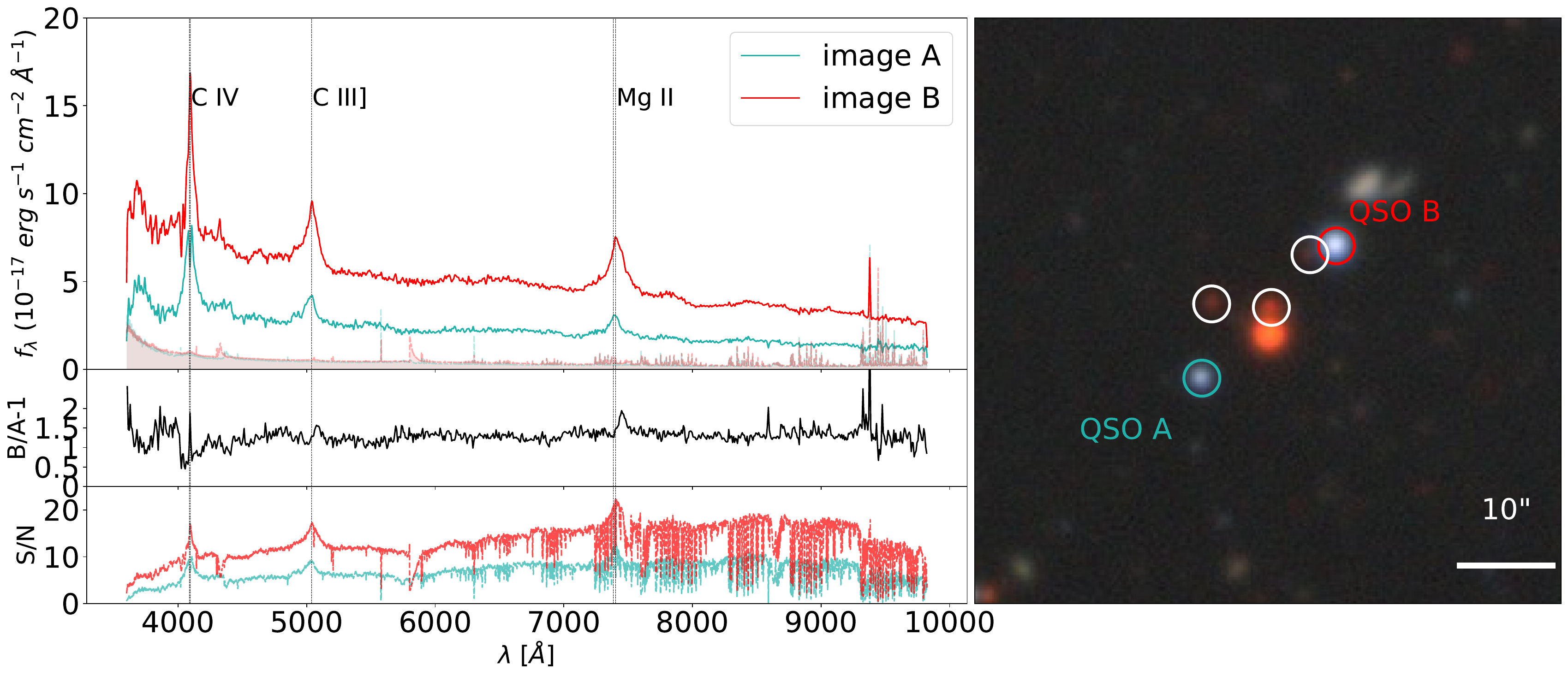}
      
    \end{minipage}
  
    \caption{\label{figure: 28001}Spectra and image of \textup{J150155.61$-$025728.4}, the image separation is $19.32\arcsec$. Left: DESI~DR1 spectra of images A and B (smoothed for clarity; shaded bands show noise); vertical markers at $z=1.646$. Right: DESI Legacy Surveys DR9 image. Data suggest that a galaxy group or cluster at \(z \approx 0.89\) may reside between the two quasar images; these plausible member galaxies are marked as white circles. See Appendix~\ref{app_section: samples with spectrumm_2} for details.}
 \end{figure*}

DESI~DR1 spectrum yields $z_A=1.6438$ and $z_B=1.6475$ (Figure \ref{figure: 28001}). The field likely hosts a foreground group/cluster about redshift $0.89$, plausibly member galaxies are marked as white circles in the right panel of Figure \ref{figure: 28001}. An SIE model centred on the $z=0.89$ galaxy fits the observed geometry, with $\sigma_v\simeq989.0~\mathrm{km\,s^{-1}}$, $q\simeq0.98$, $\phi\simeq29.2^\circ$, $\theta_{\rm E}\simeq9.36\arcsec$, and $M(<\theta_{\rm E})\simeq5.5\times10^{13}~M_\odot$. In addition, we do not find this object in the quasar pair catalogue of J25, because the projected physical distance of this object is larger than the upper limit on the projected physical distance adopted in J25. Confirmation requires deeper imaging to reveal additional lensing features. (Full details are provided in Appendix~\ref{app_section: samples with spectrumm_2}.)

\subsection{Dual quasars} \label{section: dual quasar}

After the subsample with spectroscopic information had been purged of likely WSLQ systems, to select dual quasars, we imposed two additional constraints that the maximum projected separation between group members is less than \(100\ \mathrm{kpc}\) and the velocity difference below \(2000\ \mathrm{km\,s^{-1}}\), the velocity difference criterion simultaneously accounts for the peculiar-velocity difference of the dual quasars and the uncertainty in their broad-line spectroscopic redshifts J25. Those criteria yielded a set of 29 dual quasar candidates. The complete catalogue and the DESI legacy survey/Pan-STARRS1 images of each dual quasar candidate are also available online. Table~\ref{tb:dualquasar} lists the principal properties of these systems; the electronic table also contains all parameters provided by the original \texttt{CatNorth}, which are not exhibited in Table~\ref{tb:dualquasar} for brevity.

\begin{figure*}
    \centering
    \begin{minipage}[t]{1\linewidth}
        \centering
        \includegraphics[width=1\linewidth]{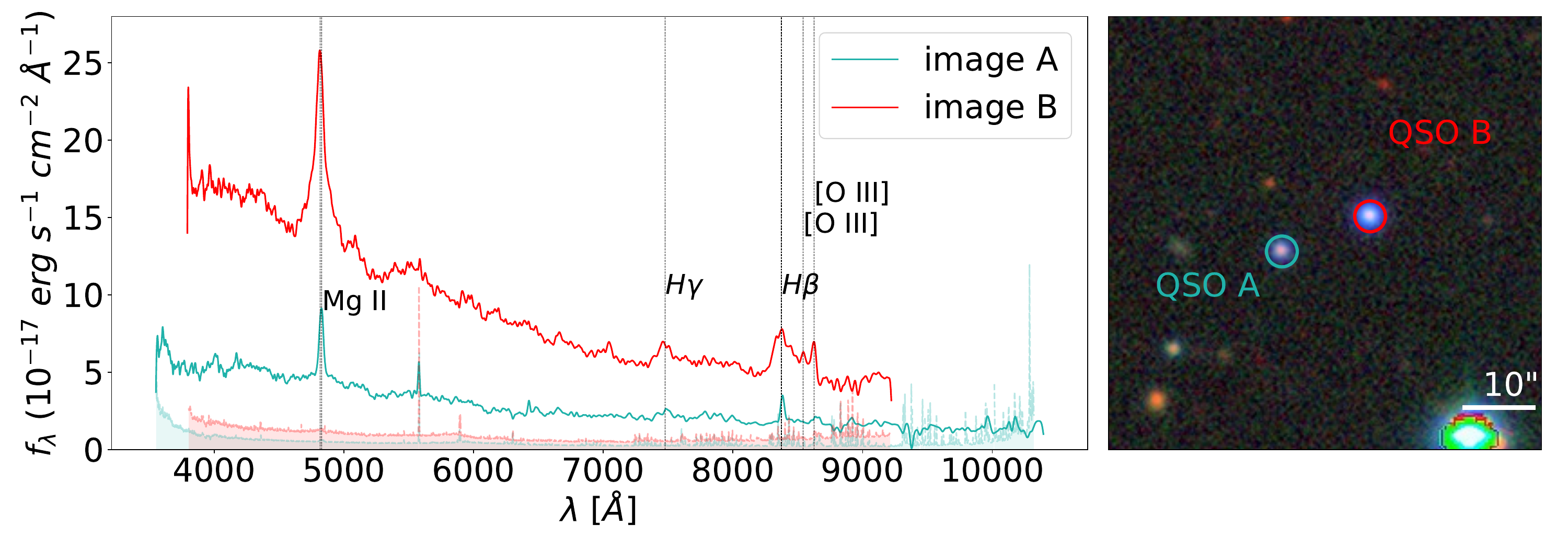}
      
    \end{minipage}
    \begin{minipage}[t]{1\linewidth}
        \centering
        \includegraphics[width=1\linewidth]{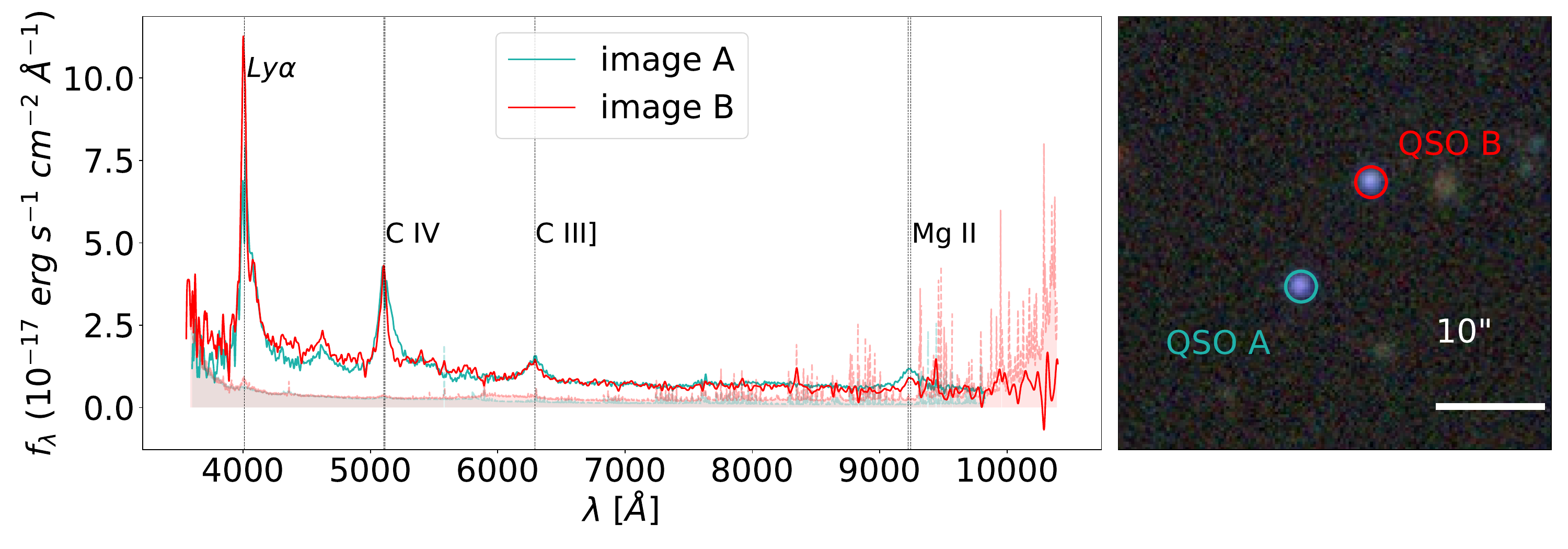}
      
    \end{minipage}
   
    \caption{\label{figure: showcase_dualquasar9333and26644}Spectra and images of two dual quasar candidates. In each row the left sub–panel shows the spectra and the right sub–panel presents the DESI Legacy Imaging Surveys DR9 \(grz\) composite image.  \textit{Top:} dual quasar candidate of \texttt{groupid}~\(=9333\).  Spectrum of image A is from SDSS DR16 and that of image B is from DESI DR1; automated spectral redshifts are \(z_{A}=0.7239\) and \(z_{B}=0.7201\).  The angular separation is 13.17 arcsec, corresponding to a projected distance of \(95.01\ \mathrm{kpc}\); the velocity difference is \(661.56\ \mathrm{km\,s^{-1}}\).  \textit{Bottom:} dual quasar candidate of \texttt{groupid}~\(=26644\).  Spectrum of image A derives from DESI DR1 and image B from SDSS DR16 with automated spectral redshifts \(z_{A}=2.3028\) and \(z_{B}=2.2938\).  The angular separation is 11.61 arcsec, giving a projected distance of \(96.51\ \mathrm{kpc}\); the velocity difference is \(818.04\ \mathrm{km\,s^{-1}}\).}
 \end{figure*}

\begin{figure*}
    \centering
    \begin{minipage}[t]{1\linewidth}
        \centering
        \includegraphics[width=1\linewidth]{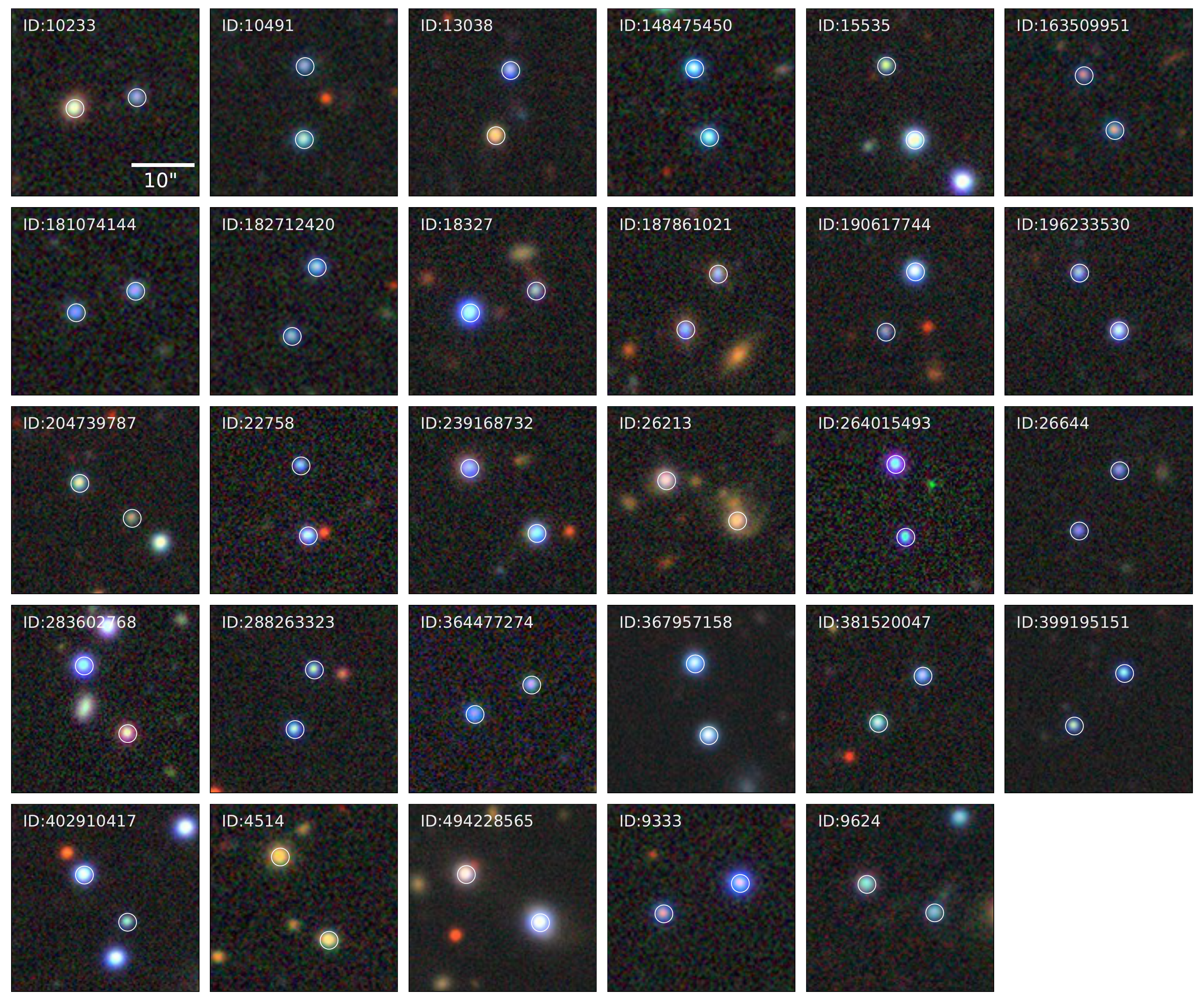}
      
    \end{minipage}
 
    \caption{\label{figure: showcase_dualquasar}DESI Legacy Imaging Surveys DR9 \(grz\) composite images of the 29 dual quasar candidate systems. Each cut-out spans 30 arcsec $\times$ 30 arcsec.  White circles mark the positions of the quasar candidate images.}
 \end{figure*}

Figure~\ref{figure: showcase_dualquasar} presents cut-outs from the DESI Legacy Imaging Survey DR9 of the 29 dual quasar candidates.  Each panel spans the same angular extent of a sidelength of 30 arcsec.  These image pairs either lack the evidence for the presence of enough foreground matter to form strong gravitational lensing or have clearly different spectral redshift or features.  Figures~\ref{figure: showcase_dualquasar9333and26644} displays the optical spectra of two randomly selected systems; the upper system and lower systems' velocity differences are \(661.56\ \mathrm{km\,s^{-1}}\) and \(818.04\ \mathrm{km\,s^{-1}}\), and their projected separations are \(95.01\ \mathrm{kpc}\) and \(96.51\ \mathrm{kpc}\), respectively.

J25 selected 1\,842 quasar pairs from the DESI~DR1 spectra.  Cross matching our 29 dual quasar candidate samples with J25 shows that 14 have counterparts in J25. The other 15 dual quasar candidates are not included in J25 because their spectra are not all from DESI DR1, i.e., the spectrum of at least one image of the system is from SDSS DR16. The final table records this information in the column labelled \texttt{in\_J25}, which flags whether the sample is present in J25.

\begin{table*}[htbp]
\caption{Column description of the dual quasar candidates catalogue.}
\label{tb:dualquasar}
\centering
\small
\begin{tabular}{p{0.04\textwidth}@{\hspace{3pt}}
                p{0.20\textwidth}@{\hspace{3pt}}
                p{0.06\textwidth}@{\hspace{3pt}}
                p{0.07\textwidth}@{\hspace{3pt}}
                p{0.55\textwidth}}
\hline\hline
Col. & Name & Type & Unit & Description\\
\hline
 1  & \texttt{ra}                   & double & deg                     & \textit{Gaia}\,DR3 right ascension (ICRS, epoch 2016.0)\\
 2  & \texttt{dec}                  & double & deg                     & \textit{Gaia}\,DR3 declination (ICRS, epoch 2016.0)\\
 3  & \texttt{groupid}            & int  & ---                     & System identifier\\
 4  & \texttt{dv}                   & float  & km\,s$^{-1}$            & Line-of-sight velocity difference\\
 5  & \texttt{sep\_max}             & float  & arcsec                  & Maximum angular separation of the pair\\
 6  & \texttt{dis}                  & float  & kpc                     & Maximum projected separation\\
 7  & \texttt{Wen\_BCG\_RA}         & float  & deg                     & R.A.\ of matched BCG in \textsc{WEN\_CAT}\\
 8  & \texttt{Wen\_BCG\_DEC}        & float  & deg                     & Dec.\ of the same BCG\\
 9  & \texttt{Wen\_redshift}        & float  & ---                     & Cluster redshift in \textsc{WEN\_CAT}\\
10  & \texttt{Wen\_M500}            & float  & $10^{14}$\,M$_\odot$    & $M_{500}$ from \textsc{WEN\_CAT} richness\\
11  & \texttt{Wen\_ID}              & int    & ---                     & Cluster identifier in \textsc{WEN\_CAT}\\[2pt]
12  & \texttt{Zou\_BCG\_RA}         & float  & deg                     & R.A.\ of matched BCG in \textsc{ZOU\_CAT}\\
13  & \texttt{Zou\_BCG\_DEC}        & float  & deg                     & Dec.\ of the same BCG\\
14  & \texttt{Zou\_redshift}        & float  & ---                     & Cluster redshift in \textsc{ZOU\_CAT}\\
15  & \texttt{Zou\_M500}            & float  & $\log_{10}$(M$_\odot$) & $M_{500}$ in \textsc{ZOU\_CAT}\\
16  & \texttt{Zou\_ID}              & int    & ---                     & Cluster identifier in \textsc{ZOU\_CAT}\\[2pt]
17  & \texttt{eROSITA\_RA}          & float  & deg                     & R.A.\ of matched eROSITA cluster\\
18  & \texttt{eROSITA\_DEC}         & float  & deg                     & Dec.\ of the same cluster\\
19  & \texttt{eROSITA\_redshift}    & float  & ---                     & eROSITA cluster redshift\\
20  & \texttt{eROSITA\_M500}        & float  & $10^{13}$\,M$_\odot$    & $M_{500}$ in eROSITA catalogue\\
21  & \texttt{eROSITA\_ID}          & int    & ---                     & Cluster identifier in eROSITA catalogue\\[2pt]
22  & \texttt{in\_J25}              & flag   & ---                     & Whether this system exists in J25\\
\hline
\end{tabular}
\end{table*}

\section{Discussion}\label{section: discussion}

This section is arranged as follows. Section~\ref{section: completeness} discusses the completeness of the WSLQ candidate sample obtained in this work. Section~\ref{section: Proportion of double lensed quasar} analyses the proportion of two image systems within that sample. Section~\ref{section: follow up} introduces planned follow-up aimed at confirming the nature of these candidates and the potential scientific significance.

\subsection{Completeness}\label{section: completeness}

The completeness of the known WSLQs in \texttt{CatNorth} is 50 per cent, four of the eight published systems are discoverable in \texttt{CatNorth}.  Of the remaining four, only COOL J0542$-$2125 \citep{Martinez_2023} has its brightest image (photometric information comes from the DESI Legacy Imaging Survey DR9, see Table \ref{tb:wslq_known}) detected in \texttt{CatNorth}, whereas SDSS J2222+2745 \citep{2013ApJ...773..146D}, SDSS J0909+4449 \citep{Shu_2018}, and COOL J0335$-$1927 \citep{napier2023coollampsvdiscoverycool} lack any counterparts in \texttt{CatNorth}.  The principal reason is that the magnitudes of their lensed images lie around or below the limiting magnitude of \texttt{CatNorth}. In addition, the performance of the four discoverable WSLQs within our candidate-selection workflow indicates that the pre-VI stages, namely the quasar group finder and the automatic screening, are consistent with a high completeness for potential lensed quasars present in \texttt{CatNorth}, albeit based on only four systems.


\subsection{Proportion of lensed quasar candidates with two images}\label{section: Proportion of double lensed quasar}

Among the 331 lensed quasar candidate systems identified in this work, only two display three images, whereas the remaining 329 have two image counterparts in \texttt{CatNorth}. This proportion of two-image systems is much higher than the proportion of doubles predicted by theoretical simulations (see, e.g., \citealt{2004ApJ...610..663O}), and is also far higher than the observed proportion of doubles in the set of eight published WSLQs, where only SDSS\,J1326\,+\,4806 is a double image system. Important reasons for the difference are the restricted limiting magnitude of \textit{Gaia}/\texttt{CatNorth} and the non-zero probability that the \texttt{CatNorth} construction process removed one or more true quasar images. These factors likewise explain why three of the four discoverable known lenses (the ones with N$\geq$3) present in \texttt{CatNorth}, i.e. J1004\,+\,4112, SDSS\,J1029\,+\,2623, and GraL\,J165105.3$-$041725, which possess four, three, and four images respectively, appear with only two image counterparts recorded in \texttt{CatNorth}. Consequently, any as-yet-unknown positive systems that can be detected in \texttt{CatNorth} are likely to be represented by no more than two images. In addition, another possible factor is that criterion (i) in VI may cause us to miss three-image systems of the naked cusp configuration in our final candidate sample, which would finally tend to decrease the fraction of N>2 image systems.

\subsection{Follow-up and scientific significance}\label{section: follow up}

We plan to carry out follow-up confirmation of the strong lensing candidates identified in this study, employing distinct verification strategies tailored to each class of candidate. For the high-quality lensed quasar candidates, especially those that can be matched with galaxy clusters in Grade-A and the samples which already have spectral information, we plan to request observing time on the Canada-France-Hawaii Telescope (CFHT) and the 200-inch Hale Telescope (P200) to follow them up. For the remaining lensed quasar candidates, we will cross-match with DESI future release spectroscopy, and - as public releases expand - Euclid imaging/slitless spectroscopy, future survey observations on the 4-metre Multi-Object Spectroscopic Telescope (4MOST) and the forthcoming CSST imaging/slitless spectroscopy wide-field surveys when data is available.

If confirmed, these candidates would offer substantial scientific value. For instance, within the subset where the BCG of a galaxy cluster lies within 30 arcsec of the quasar-group centre (shown in  Figure~\ref{figure: showcases of wo spec sample with BCG or ero}), the rightmost panel of Figure~\ref{figure: showcases of wo spec sample with BCG or ero} shows the candidate with \texttt{groupid} \(=37148\), whose image separation is 37.59 arcsec, exceeding that of all currently known WSLQs. If genuine, its large separation would provide a uniquely valuable case for probing the three-dimensional structure of quasars \citep{Misawa_2016}. As another example, Appendix~\ref{app_section: samples with spectrumm_1} discusses a candidate in which the mass centroid and luminosity centroid may be offset; if verified, this system would be highly informative for constraining the mass distribution of this irregular foreground galaxy cluster.

\section{Summary}\label{section: conclusion}

In this work, the catalogue-based searching strategy of \citet{2023A&A...672A.123H} is applied to \texttt{CatNorth}, a high purity and highly complete quasar candidate catalogue derived from \textit{Gaia}~DR3, to search for WSLQ candidates. The analysis delivers and releases two samples: a set of WSLQ candidates and a set of dual quasar candidates. The procedure consists of three steps.  
(i) The group finder is run on the 1\,545\,514 quasars and generates about 24\,000 quasar groups.  
(ii) These groups are filtered by their colour properties or by spectra retrieved from SDSS DR16 and DESI DR1, this operation reduces the sample to about 14\,000 groups.  
(iii) Visual inspection assigns scores and categories to the surviving systems and removes those that are very likely to be spurious.

The resulting WSLQ candidate sample contains two systems for which spectroscopic data are available and 331 systems selected solely on the basis of colour and imaging information in the absence of adequate spectroscopy.  The two systems with spectroscopic records have been subjected to spectral analysis and preliminary lens modelling. The 331 colour-selected candidates were scored and placed into three classes: 45 Grade-A, 98 Grade-B, and 188 Grade-C.  The complete sample has been cross matched with roughly 1.9 million galaxy clusters and groups. 108 samples successfully matched at least one galaxy cluster within 2 arcmin in three cluster catalogues which we used, and the result is included in the publicly available table.

Applying the additional criteria of a velocity difference \(\Delta v < 2000\ \mathrm{km\,s^{-1}}\) and a projected separation smaller than \(100\ \mathrm{kpc}\) on the remaining quasar candidate groups whose spectra are available yields a catalogue of 29 dual quasar candidates. Their images show no morphology or spectral features characteristic of strong lensing, but their small line-of-sight velocity separations suggest that they satisfy conventional criteria for the dual quasar.

We plan to obtain supplementary spectroscopy and deep imaging for the candidates identified in this study using CFHT, P200, and DESI future release, while the ongoing and forthcoming wide-field surveys with Euclid and CSST will aid the confirmation of lensed quasars in the long term. Spectroscopic observations will refine the sample by confirming plausible strong lens systems and discarding obvious contaminants, whereas deeper imaging will reveal additional strong lensing indicators and impose tighter constraints on the mass distribution of the foreground deflectors. Together, these data are essential for establishing the physical nature of the candidates.

In summary, we have implemented a robust pipeline that starts from the quasar candidate catalogue \texttt{CatNorth} and produces a catalogue of WSLQ candidates together with a catalogue of dual quasar candidates; lens modelling has been carried out for the two high-quality systems whose spectroscopic information is available. Follow-up spectroscopy and deep imaging of the most promising candidates are planned to confirm new WSLQs based on these samples in the near future. And, as part of our ClUsteR strong Lens modellIng for the Next-Generation observations (CURLING) program \citep{xie2024curlingiinfluence,Xie_2025}, the newly confirmed WSLQs will be used to constrain the properties of dark matter and dark energy by employing our advanced pixelized lens-modelling technique. This study also demonstrates that the proposed automatic procedure can markedly compress a very large parent sample of quasar candidates while maintaining high completeness, an ability that will be invaluable for efficiently selecting WSLQs in the next-generation deep, wide-field imaging surveys.


\begin{acknowledgements}

We thank astropy, HEALPix, pandas, and lenstronomy for providing convenient and reliable Python packages. We thank Yiping Shu for insightful discussions. Z.H. acknowledges support from the National Natural Science Foundation of China (Grant No. 12403104). N.L. acknowledges the support of the science research grants from the China Manned Space Project (No. CMS-CSST-2021-A01) and the CAS Project for Young Scientists in Basic Research (No. YSBR-062).

The Pan-STARRS1 Surveys (PS1) and the PS1 public science archive have been made possible through contributions by the Institute for Astronomy, the University of Hawaii, the Pan-STARRS Project Office, the Max-Planck Society and its participating institutes, the Max Planck Institute for Astronomy, Heidelberg and the Max Planck Institute for Extraterrestrial Physics, Garching, The Johns Hopkins University, Durham University, the University of Edinburgh, the Queen's University Belfast, the Harvard-Smithsonian Center for Astrophysics, the Las Cumbres Observatory Global Telescope Network Incorporated, the National Central University of Taiwan, the Space Telescope Science Institute, the National Aeronautics and Space Administration under Grant No. NNX08AR22G issued through the Planetary Science Division of the NASA Science Mission Directorate, the National Science Foundation Grant No. AST–1238877, the University of Maryland, Eotvos Lorand University (ELTE), the Los Alamos National Laboratory, and the Gordon and Betty Moore Foundation.

This research used data obtained with the Dark Energy Spectroscopic Instrument (DESI). DESI construction and operations is managed by the Lawrence Berkeley National Laboratory. This material is based upon work supported by the U.S. Department of Energy, Office of Science, Office of High-Energy Physics, under Contract No. DE–AC02–05CH11231, and by the National Energy Research Scientific Computing Center, a DOE Office of Science User Facility under the same contract. Additional support for DESI was provided by the U.S. National Science Foundation (NSF), Division of Astronomical Sciences under Contract No. AST-0950945 to the NSF’s National Optical-Infrared Astronomy Research Laboratory; the Science and Technology Facilities Council of the United Kingdom; the Gordon and Betty Moore Foundation; the Heising-Simons Foundation; the French Alternative Energies and Atomic Energy Commission (CEA); the National Council of Humanities, Science and Technology of Mexico (CONAHCYT); the Ministry of Science and Innovation of Spain (MICINN), and by the DESI Member Institutions: www.desi.lbl.gov/collaborating-institutions. The DESI collaboration is honored to be permitted to conduct scientific research on I’oligam Du’ag (Kitt Peak), a mountain with particular significance to the Tohono O’odham Nation. Any opinions, findings, and conclusions or recommendations expressed in this material are those of the author(s) and do not necessarily reflect the views of the U.S. National Science Foundation, the U.S. Department of Energy, or any of the listed funding agencies.

The DESI Legacy Imaging Surveys consist of three individual and complementary projects: the Dark Energy Camera Legacy Survey (DECaLS), the Beijing-Arizona Sky Survey (BASS), and the Mayall z-band Legacy Survey (MzLS). DECaLS, BASS and MzLS together include data obtained, respectively, at the Blanco telescope, Cerro Tololo Inter-American Observatory, NSF’s NOIRLab; the Bok telescope, Steward Observatory, University of Arizona; and the Mayall telescope, Kitt Peak National Observatory, NOIRLab. NOIRLab is operated by the Association of Universities for Research in Astronomy (AURA) under a cooperative agreement with the National Science Foundation. Pipeline processing and analyses of the data were supported by NOIRLab and the Lawrence Berkeley National Laboratory (LBNL). Legacy Surveys also uses data products from the Near-Earth Object Wide-field Infrared Survey Explorer (NEOWISE), a project of the Jet Propulsion Laboratory/California Institute of Technology, funded by the National Aeronautics and Space Administration. Legacy Surveys was supported by: the Director, Office of Science, Office of High Energy Physics of the U.S. Department of Energy; the National Energy Research Scientific Computing Center, a DOE Office of Science User Facility; the U.S. National Science Foundation, Division of Astronomical Sciences; the National Astronomical Observatories of China, the Chinese Academy of Sciences and the Chinese National Natural Science Foundation. LBNL is managed by the Regents of the University of California under contract to the U.S. Department of Energy.

This publication makes use of data products from the Wide-field Infrared Survey Explorer, which is a joint project of the University of California, Los Angeles, and the Jet Propulsion Laboratory/California Institute of Technology, funded by the National Aeronautics and Space Administration.

This work has made use of data from the European Space Agency (ESA) mission Gaia (https://www.cosmos.esa.int/gaia), processed by the Gaia Data Processing and Analysis Consortium (DPAC, https://www.cosmos.esa.int/web/gaia/dpac/consortium). Funding for the DPAC has been provided by national institutions, in particular the institutions participating in the Gaia Multilateral Agreement.

Funding for the Sloan Digital Sky 
Survey IV has been provided by the 
Alfred P. Sloan Foundation, the U.S. 
Department of Energy Office of 
Science, and the Participating 
Institutions.

SDSS-IV acknowledges support and 
resources from the Center for High 
Performance Computing  at the 
University of Utah. The SDSS 
website is www.sdss4.org.

SDSS-IV is managed by the 
Astrophysical Research Consortium 
for the Participating Institutions 
of the SDSS Collaboration including 
the Brazilian Participation Group, 
the Carnegie Institution for Science, 
Carnegie Mellon University, Center for 
Astrophysics | Harvard \& 
Smithsonian, the Chilean Participation 
Group, the French Participation Group, 
Instituto de Astrof\'isica de 
Canarias, The Johns Hopkins 
University, Kavli Institute for the 
Physics and Mathematics of the 
Universe (IPMU) / University of 
Tokyo, the Korean Participation Group, 
Lawrence Berkeley National Laboratory, 
Leibniz Institut f\"ur Astrophysik 
Potsdam (AIP),  Max-Planck-Institut 
f\"ur Astronomie (MPIA Heidelberg), 
Max-Planck-Institut f\"ur 
Astrophysik (MPA Garching), 
Max-Planck-Institut f\"ur 
Extraterrestrische Physik (MPE), 
National Astronomical Observatories of 
China, New Mexico State University, 
New York University, University of 
Notre Dame, Observat\'ario 
Nacional / MCTI, The Ohio State 
University, Pennsylvania State 
University, Shanghai 
Astronomical Observatory, United 
Kingdom Participation Group, 
Universidad Nacional Aut\'onoma 
de M\'exico, University of Arizona, 
University of Colorado Boulder, 
University of Oxford, University of 
Portsmouth, University of Utah, 
University of Virginia, University 
of Washington, University of 
Wisconsin, Vanderbilt University, 
and Yale University.

\end{acknowledgements}

\bibliographystyle{aa}
\bibliography{cite}

\clearpage
\begin{appendix}

\section{Candidates with spectrum}\label{section: appendix}

Spectroscopic observations of quasars are invaluable for excluding systems that are not real strong lenses. Microlensing \citep{2012A&A...544A..62S, Motta2012, Hutsem_kers_2023}, variability in the broad emission lines and continuum \citep{2019ApJS..241...34S}, differential reddening, differences in sight-line \citep{Misawa_2016} can introduce differences between the spectra of multiple images. Nevertheless, the spectra of confirmed strong-lens systems are, in an overall sense, highly similar, and any systematic discrepancies can be attributed to the effects listed above. 


This section discusses two systems with spectroscopical information that are plausible strong lenses and presents their lens modelling process. A summary of the properties of these two samples is provided in Section \ref{section:spectro_cands} of the main text. Section~\ref{section: lens modeling} outlines the modelling methodology, whereas Section~\ref{section: samples with spectrumm} describes the information of two candidate strongly lensed quasars and the corresponding modelling results.

\subsection{Lens modelling methodology}\label{section: lens modeling}

In our lens modelling methodology the foreground lens is represented with a singular isothermal ellipsoid (SIE) mass profile. The model contains three free parameters: the velocity dispersion $\sigma_{v}$, the axis ratio $q$, and the position angle $\phi$ of the major axis of the mass contour. Deflection angles and image positions are computed with the \texttt{lenstronomy} package \citep{2018PDU....22..189B,2021JOSS....6.3283B}, and the parameter posterior is sampled by the Markov chain Monte Carlo ensemble sampler \texttt{emcee} \citep{Foreman_Mackey_2013}. The likelihood is built from the positional offsets between the model quasar image and the observations,  
\begin{equation}
\chi^{2}_{\mathrm{pos}}
=\sum_{i} \frac{\lvert \vec{r}^{\,M}_{i}-\vec{r}_{i} \rvert^{2}}{\sigma_{i}^{2}},
\label{eq:chi2_position}
\end{equation}
where $\vec{r}^{\,M}_{i}$ and $\vec{r}_{i}$ are the modelled and observed positions of the $i$th image, and $\sigma_{i}$ is the corresponding astrometric uncertainty.

The SIE profile offers the advantage of a small parameter set, which limits degeneracies when only a few image positions are available. Its simplicity, however, makes it an imperfect description of cluster scale haloes, whose central density profiles are usually flatter and often require multiple subhaloes for an accurate modelling \citep{2020ApJS..247...12S}. But given the limited observational constraints presently available for the candidates discussed here, the SIE model provides a useful first approximation; more elaborate mass models should be adopted once deeper imaging and additional spectroscopy become accessible.

\subsection{Samples} \label{section: samples with spectrumm}

\subsubsection{\textup{J110121.67+060931.3}} \label{app_section: samples with spectrumm_1}

Two quasar images in this system have similar spectral features, and a foreground galaxy cluster acts as a plausible deflector. The spectra and DESI Legacy Imaging Surveys DR9 image of this system are displayed in Figure~\ref{figure: 359463625}, left and right panels respectively.  Images~A and~B are separated by 14.14\ arcsec. In the uppermost panel of the spectral figure, the spectrum is convolved with a scale kernel whose width is five pixels for smoothing. Both spectra originate from DESI-DR1, automatic DESI-DR1 redshift fits yield \(z_{A}=0.8313 \pm 0.0001\) and \(z_{B}=0.8297 \pm 0.0002\). The spectrum of image A is of lower quality and exhibits numerous spurious “absorption” features with zero signal-to-noise (for example near \(5907\) Å), introduced when zero-filled regions were convolved with neighbouring valid data.  Image B appears slightly redder than image A, and the continuum flux ratio varies with wavelength; these differences may arise from differential reddening, microlensing, and intervening line-of-sight structure.

A foreground galaxy cluster is matched in \textsc{WEN\_CAT}, and an X-ray signal is recorded in \textit{eRASS1 Main catalogue} \citep{2024A&A...682A..34M}, which is the possible X-ray emission from the hot gas of this galaxy cluster. The BCG is marked by the white circle north of the quasar images in Figure~\ref{figure: 359463625}.  \textsc{WEN\_CAT} lists a photometric redshift \(z_{\mathrm{ph}}=0.2503\) and \(M_{500}=5.0\times10^{13}\,M_{\odot}\) of this cluster. A recorded X-ray signal from the \textit{eRASS1 Main catalogue} \citep{2024A&A...682A..34M} is located at the yellow triangle in Figure~\ref{figure: 359463625}.  The position uncertainty of this X-ray signal in \textit{eRASS1 Main catalogue} is 2.41 arcsec; the offset between this X-ray signal and image A和image B is 4.264 arcsec, so the X-ray emission may also originate from quasar image A.  The 0.2 - 2.3 keV flux of this X-ray signal is \(1.86\times10^{-13}\,\mathrm{erg\,s^{-1}\,cm^{-2}}\). Adopting this value as a lower limit for the galaxy cluster, with the assumption that the redshift equals 0.2503, yields a luminosity lower limit of \(2.95\times10^{43}\,\mathrm{erg\,s^{-1}}\), using the empirical mass luminosity relation of \citet{2002ApJ...567..716R} for the 0.1 - 2.4 keV band gives \(M_{500}\gtrsim2\times10^{14}\,M_{\odot}\).\footnote{K-correction is neglected here; for galaxy clusters with kT=0.5 - 10 keV at \(z=0.25\), the K-correction affects the luminosity by 0 to about 13 per cent \citep{2004A&A...425..367B}.}

Lens modelling of this system follows the methodology outlined in Section~\ref{section: lens modeling}. Two centring hypotheses are examined. In the first, the BCG is adopted as the centre of the mass model; under this assumption a configuration with only two quasar images cannot be reconciled with the observations, and a search of the DESI Legacy Imaging Surveys DR9 image reveals no additional possible quasar images with similar colour in the regions where three- or four-image configurations would place them. In the second hypothesis the centre of the mass distribution is taken to coincide with the peak of the X-ray emission. Under this assumption the observed quasar images are well reproduced by the SIE model. The best fit gives \(\sigma_{v}=608.58\ \mathrm{km\,s^{-1}}\), \(q=0.89\), and \(\phi=19.32^{\circ}\). The left panel of Figure~\ref{figure: lens modelling} displays the corresponding lens model.

Following the galaxy cluster mass estimation framework used in \cite{Shu_2019}, we estimate the mass of the second lens model assumption as follows, and found that the mass inferred based on X-ray observation is compatible with the lens model and is sufficient to generate the observed lensed quasar images separation. The Einstein radius of this lens model is 7.187 arcsec, which implies an enclosed mass of \(7.394\times10^{12}\ M_{\odot}\) within Einstein radius. For an NFW halo with a lower mass limit of \(2\times10^{14}\ M_{\odot}\) (given by X-ray observation in this sample) and concentration \(3 \le c \le 8\), the projected mass within \(R<7.187\) arcsec has a lower mass limit which lies in the range \((2.31\text{--}5.33)\times10^{12}\ M_{\odot}\), consistent with the value derived from the lens model. If the estimation \(M_{500}=5\times10^{13}\ M_{\odot}\) provided by \textsc{WEN\_CAT} is adopted, where the \(M_{500}\) is given by the richness, the projected mass inside the same radius is insufficient to produce an Einstein radius as large as 7.187 arcsec.

We want to note that all eight known WSLQs can be satisfactorily fit with cluster mass centroids that nearly coincide with the BCG. Nevertheless, a non-negligible displacement between the cluster centre of mass and the BCG is still permitted by the data, leaving the second lens modelling hypothesis (assuming the X-ray peak as mass centre) possible. Some studies find significant offsets between the hot gas peak and the central galaxy in a fraction of clusters \citep{2019MNRAS.487.2578Z,2014ApJ...797...82L,2013ApJ...767...38S,2023AAS...24146024D}. For example, \citet{2014ApJ...797...82L} find that in the sample of 433 BCGs with redshift smaller than \(0.08\) about fifteen per cent have an X\,ray-BCG offset exceeding \(100\ \mathrm{kpc}\). \citet{2023AAS...24146024D} show that among 186 clusters with \(0.1 \le z \le 1.4\), approximately 25 per cent have a central galaxy-SZ offset larger than \(330\ \mathrm{kpc}\), although refining the BCG identification reduces this fraction to about ten per cent. The same study reports that within a radius of \(0.3\) to \(1\ \mathrm{Mpc}\) clusters with small X ray-BCG offsets exhibit considerably stronger lensing signals than clusters with large offsets. These results indicate that the central galaxy is not always a reliable indicator of the gravitational potential centre. For J110121.67+060931.3, if we assume that the X-ray emission originates from the foreground galaxy cluster at redshift 0.2503 given by \textsc{WEN\_CAT}, then the offset between the X-ray position and the BCG is $47.79\,\mathrm{kpc}$; this value is within the observed range of X ray-BCG offset reported in the aforementioned literature.

We want to point out that attributing this X-ray emission to the diffuse emission of the galaxy cluster provided by WEN\_CAT is highly dubious; this model is only a suspicion. There are two reasons. First, this X-ray emission is classified as a `point-like' source (EXT\_LIKE$=0$) in the eRASS1 Main catalogue \citep{2024A&A...682A..34M}, which weakens the possibility that this X-ray emission is diffuse emission from a galaxy cluster. Another reason is that the aforementioned inferred $M_{500}\gtrsim2\times10^{14}\,M_{\odot}$ based on the $L_{\mathrm{X}}$-$M_{500}$ scaling relation and the cluster redshift of about 0.25 imply that such galaxy cluster falls well within the detection limit of the eRASS1 Galaxy Groups and Clusters catalogue \citep{bulbul2024srgerositaallskysurveycatalog}. However, this source is not present in the eRASS1 Galaxy Groups and Clusters catalogue. It supports the smaller cluster-mass inference given by WEN\_CAT, thereby weakening the possibility that this system is a WSLQ.

Further confirmation or refutation of this candidate requires deeper multi-wavelength imaging to pinpoint the cluster mass centroid and profile more accurately or to reveal fainter quasar image candidates below the DESI Legacy Imaging Surveys DR9 detection limit that would support a model centred on the BCG or another location.

      
   

\subsubsection{\textup{J150155.61-025728.4}}
\label{app_section: samples with spectrumm_2}

Figure~\ref{figure: 28001} shows the spectra (left panel) and the DESI Legacy Imaging Surveys DR9 image (right panel) of this system. The separation between images A and B is \(19.32\) arcsec. Both spectra are from DESI DR1, which yields automatic redshifts \(z_{A}=1.6438 \pm 0.0004\) and \(z_{B}=1.6475 \pm 0.0002\).

The DESI Legacy Imaging Surveys DR9 image shows that a galaxy cluster or group possibly existed between images A and B. The brightest object situated between the two quasar images is a star. A galaxy immediately to its north has a spectroscopic redshift of \(0.89\) in DESI DR1. Two fainter red objects located west and northeast (close to image B) of this galaxy have photometric redshifts of \(0.921 \pm 0.075\) and \(0.923 \pm 0.17\) in the DESI Legacy Imaging Surveys DR9 catalogue. These three images are indicated by white circles in Figure~\ref{figure: 28001}. These measurements suggest that a galaxy group or cluster at \(z \simeq 0.89\) may reside in the central region and provide the gravitational potential necessary to produce the double quasar images. Deeper imaging will be required to confirm or rule out this possibility.

The spectra of images A and B have some different features, which do not rule out the possibility of this system being a real lensing system, because these might be attributed to several mechanisms. For example, the ratio of the red to blue wings of the Mg II and C III] lines in image B differs markedly from that in image A. Such a difference could arise from microlensing \citep{2012A&A...544A..62S, Motta2012, Hutsem_kers_2023}. At the position of C IV in image A, and on the blue wing of C IV in image B, multiple different narrow absorption lines (NALs) are present; these could be caused by the difference in quasar inner structure at different sightlines of the two images \citep{Misawa_2016}. An alternative possibility is that some of these narrow absorption features originate in intervening material that is unrelated to the quasar.

\begin{figure*}
    \centering
    \begin{minipage}[t]{0.48\linewidth}
        \centering
        \includegraphics[width=1\linewidth]{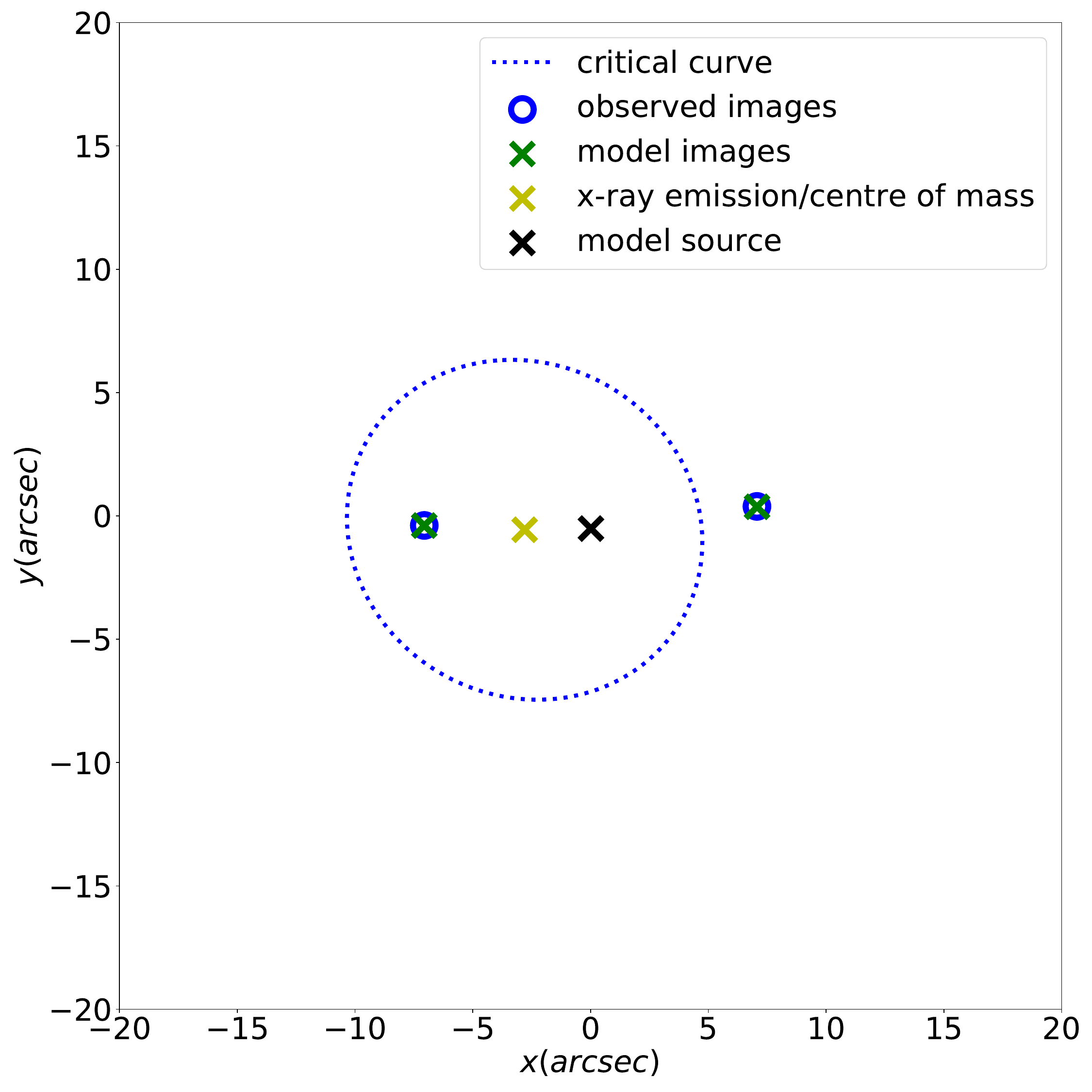}
      
    \end{minipage}
    \begin{minipage}[t]{0.48\linewidth}
        \centering
        \includegraphics[width=1\linewidth]{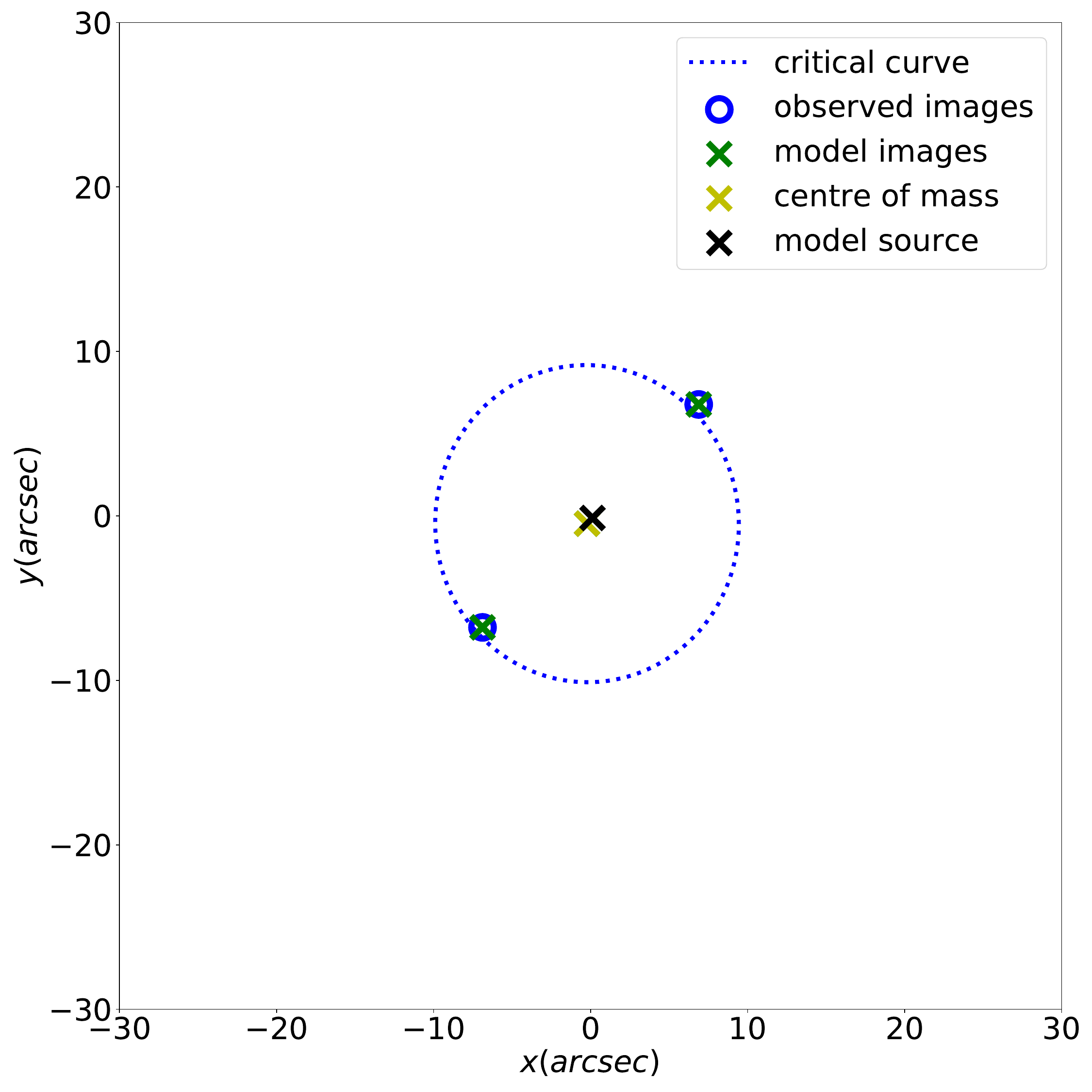}
    
    \end{minipage}

    
    \caption{\label{figure: lens modelling} 
    Left panel: Lens model for \textup{J110121.67+060931.3}, the system shown in Figure~\ref{figure: 359463625}, assuming that the lens centre is fixed at the plausible X-ray emission peak. Right panel: Lens model solution \textup{J150155.61$-$025728.4}, the system shown in Figure~\ref{figure: 28001}, assuming that the lens centre is fixed at the galaxy north of the bright central star whose spectroscopic redshift is \(z = 0.89\).}
    
\end{figure*}

Applying the lens modelling method outlined in Section~\ref{section: lens modeling}, the system was successfully modelled with a SIE mass profile. The lens centre was fixed at the object with spectroscopic redshift \(z = 0.89\) located north of the bright star. The model reproduces the observed configuration well: the best-fitting value gives \(\sigma_{v} = 988.96\ \mathrm{km\;s^{-1}}\), \(q = 0.98\), \(\phi = 29.19^{\circ}\), an Einstein radius of 9.364 arcsec, and a projected mass within that radius of \(5.503 \times 10^{13}\ M_{\odot}\). The best-fitting result is shown in the right-hand panel of Figure~\ref{figure: lens modelling}.

Confirmation or refutation of this candidate demands deeper imaging and additional high quality spectroscopy. Deeper data may uncover neighbouring strong-lensing features, allowing a more reliable estimate of the lens galaxy cluster properties, particularly its mass. Multiple spectroscopic observations of the quasar images will help to determine whether the observed spectral differences are produced by microlensing, differential absorption, or other mechanisms.

      
  

\section{Simulating multi-image maximum separation distribution}\label{section: appendixB}

We employed Monte Carlo simulation to predict the distribution of the maximum image angular separation of lensed quasars produced by galaxy cluster lenses, the relevant result is shown in Figure \ref{figure: wospectrum_stats}. The simulation comprises three parts: a foreground light cone, a background light cone, and the lensing simulation.

For the deflectors in the foreground light cone we adopt the cluster redshifts, masses, and sky coordinates provided by the ZOU\_CAT, but assign halo ellipticities through a Monte Carlo procedure. This procedure incorporates redshift evolution:  we set the ellipticity distribution at $z = 0$ to have a mean $\overline{e} = 0.33$ \citep{2005ApJ...618....1H} and standard deviation $\sigma_{e} = 0.16$ \citep{1991MNRAS.249..662P}, and let the mean evolve as $\overline{e} = 0.33 + 0.05z$ \citep{2005ApJ...618....1H}, and assume $\sigma_{e}$ does not evolve along redshift \citep{2006MNRAS.367.1781A,Suto_2016}; the position angle of the major axis is drawn from a uniform distribution. We assume two alternative mass models for the foreground haloes, SIE and eNFW; for the latter, the concentration parameter is computed from the fitting formulae of \cite{2018ApJ...859...55C}, while the nonlinear (“collapse”) mass scale $M_{\star}$ is calculated with the public Python package \texttt{COLOSSUS} \citep{2018ApJS..239...35D}.

The background quasar population is likewise generated by Monte Carlo sampling: source redshifts follow the quasar redshift distribution of \texttt{CatNorth}, whereas their sky positions are assumed to be uniformly distributed.

Based on the foreground light cone and the background quasar population, we perform the lensing simulation to generate the multi-image separation distribution. During the lensing simulation, mock quasars are treated as point sources. For each of the sources, we include all foreground haloes within $100^{\prime\prime}$ of the quasar and use the \texttt{lenstronomy} package \citep{2018PDU....22..189B,2021JOSS....6.3283B} to solve the maximum image separation angle of the lensed quasar (when multiple images are produced).

\end{appendix}

\end{document}